\documentclass[manuscript]{aastex}

\long\def\symbolfootnote[#1]#2{\begingroup%
\def\thefootnote{\fnsymbol{footnote}}\footnote[#1]{#2}\endgroup}

\newcommand{\capta}{the magnetic towers in the adiabatic (left), the rotating (middle) and 
the cooling (right) cases. From top to bottom time is equal to 42,~84~and
118\,yr}

\shorttitle{Structure and stability of magnetic tower jets}
\shortauthors{Huarte-Espinosa, Frank \& Blackman}

\usepackage{subfigure}
\graphicspath{{fig/}}

\begin{document}


\title{On the structure and stability of magnetic tower jets}

\author{M. Huarte-Espinosa\altaffilmark{1}, A.~Frank\altaffilmark{1}, E.~G.~Blackman\altaffilmark{1},
A.~Ciardi\altaffilmark{2,3}, P.~Hartigan\altaffilmark{4}, S.~V.~Lebedev\altaffilmark{5} and
J.~P.~Chittenden\altaffilmark{5}}

\altaffiltext{1}{Department of Physics and Astronomy, University
of Rochester, 600 Wilson Boulevard, Rochester, NY, 14627-0171}
\altaffiltext{2}{LERMA, Universit\'e Pierre et Marie Curie,
Observatoire de Paris, Meudon, France} \altaffiltext{3}{\'Ecole
Normale Sup\'erieure, Paris, France. UMR 8112 CNRS} \altaffiltext{4}{Rice
University, Department of Physics and Astronomy, 6100 S. Main,
Houston, TX 77521-1892} \altaffiltext{5}{The Blackett Laboratory,
Imperial College London, SW7 2BW London, UK}

\begin{abstract}
Modern theoretical models of astrophysical jets combine accretion,
rotation, and magnetic fields to launch and collimate supersonic
flows from a central source.  Near the source, magnetic field
strengths must be large enough to collimate the jet requiring that
the Poynting flux exceeds the kinetic-energy flux.  The extent to
which the Poynting flux dominates kinetic energy flux at large distances
from the engine distinguishes two classes of models.  In
magneto-centrifugal launch (MCL) models, magnetic fields dominate
only at scales $\lesssim 100$ engine radii, after which the jets
become hydrodynamically dominated (HD). By contrast, in Poynting
flux dominated (PFD) magnetic tower models, the field dominates
even out to much larger scales.  To compare the large distance
propagation differences of these two paradigms, we perform  3-D
ideal MHD AMR simulations of both HD and PFD 
stellar
jets formed via
the same energy flux. We also compare how thermal
energy losses and rotation of the  jet base affects the stability in
these jets. For the conditions described, we show that PFD  and HD
exhibit observationally distinguishable features:  PFD jets are
lighter, slower, and less stable than HD jets.  Unlike HD jets, PFD
jets develop current-driven instabilities  that are exacerbated as
cooling and rotation   increase, resulting in jets that are clumpier
than those in the HD limit.   Our  PFD jet simulations also resemble
the   magnetic towers that  have been recently created in  laboratory
astrophysical jet experiments.
\end{abstract}


\section{Introduction}

Non-relativistic jets are observed in the vicinities of many
Protostellar Objects, Young Stellar Objects (YSOs) and post-AGB
stars.  Plausible models suggest that jets are launched and collimated
by a symbiosis of accretion, rotation and magnetic mechanisms, which
occur at the jet ``central engine'' (see \citealp{pudritz07}, for
a review).  The jet material must be accelerated to velocities
beyond the escape speed and magnetically mediated launch models are
often favoured
because they can provide the needed directed momentum (see Livio
2004, Pudritz 2004 for reviews).

Astrophysical jets are expected to be Poynting flux dominated (PFD)
close to their engine. It is however unclear how far from the
launch region the Poynting flux continues to dominate over kinetic energy
flux, or whether the jets eventually become essentially hydrodynamic
\citep{blackman07}.  The difference between these two possibilities
is a difference between two magnetically launched outflow classes:
(1) magnetocentrifugal jets (\citealp{blandford82}; \citealp{ouyed97};
\citealp{blackman01}; \citealp{mohamed07}), in which magnetic fields
only dominate out to the Alfv\'en radius, or  (2) Poynting flux
dominated magnetic tower jets (\citealp{shibata86}; \citealp{bell96}; \citealp{ustyugova00};
\citealp{lovelace02}; \citealp{nakamura04}) in which magnetic fields
dominate the jet structure, acting as a magnetic piston over very
large distances from the engine. Indeed, magnetic fields with
initially poloidal (radial and vertical) dominant geometries anchored
to accretion discs have been shown to form tall, highly wound and
helical magnetic structures, or magnetic towers, that expand
vertically when laterally supported in pressure equilibrium with
the ambient gas (\citealp{shibata86,bell96,bell03}).

PFD jets carry large electric currents which generate strong, tightly
wound helical magnetic fields around the jet axis.  Simulations of
such jets have found that magnetic fields play a role in the formation
of current-driven kink instabilities and the stabilization of
Kelvin-Helmholtz (KH) modes (e.g. see \citealp{nakamura04}).
However, while the correlation between the mechanical power of astrophysical
jets and their main observable features (e.g. length, velocity,
cocoon geometry, etc.) has been widely studied for kinetic-energy dominated jets, this is not the
case for PFD  magnetic tower jets.

Recently, magnetized jets have been produced in laboratory experiments.
These flows appear to exhibit key aspects of magnetic tower evolution
(\citealp{lab1,suzuki}).  In these experiments, performed on Pulsed
Power current generators, the local injection of purely toroidal
magnetic energy produced high Mach number ($\sim\,$20), fully
radiative and fully magnetized jets. These magnetic towers exhibit 
Poynting flux dominated cavities with $\beta<1$ (where $\beta$ is the ratio of thermal
to magnetic pressures) which expand supersonically into an unmagnetized
ambient medium.  Within the cavity, a central jet forms via hoop
stresses.  While the body of these jets has $\beta < 1$, their core
is a high $\beta$, kinetic energy dominated flow.  The central jet
evolution of these experiments also showed the growth of current-driven,
non-linear instabilities, in particular the kink, $m=\,$1, mode.
As a result, the laboratory jets are eventually corrugated and
become a collimated chain of magnetized ``clumps'' \citep{lab1}.
These experiments were then modeled via resistive MHD simulations
specifically developed for laboratory studies \citep{lab2}, where
the details of the flow, including current distributions, were
followed. The break-up of the jet into a sequence of collimated
clumps has been suggested as an explanation for clumpy flows observed
in YSO outflow systems (Hartigan \& Morse 2007; Yirak et al. 2010;
Hartigan et al. 2011).

We note that the effect of plasma cooling via optically thin
radiation has not been followed before in simulations of
magnetic tower jets. Studies of weakly magnetized, kinetic
energy-dominated jets show that this type of cooling can make the flow more susceptible
to instabilities, such as KH modes (\citealp{hardee97}, and references
therein). 
%
Recently, 
\citet{ohsuga09} studied magnetocentrifugally launched jets with
2D radiation-MHD simulations. Ohsuga et al. found that the strength of
radiative cooling, which they control by changing the plasma density, 
affects the structure and evolution of both accretion disks and their
associate jets. Although we follow thermal energy losses in the present study, 
we do not compute radiate transfer.
%

In this paper we study the effects that thermal energy losses and
rotation, independently of one another, have on the stability of PFD
magnetic towers.  For comparison we also run simulations of collimated
asymptotically HD jets.  Such HD jets could represent the asymptotic
propagation regimes of magneto-centrifugally launched flows, which
are distinct from PFD ones in that the latter remain dominated by magnetic
flux out to much larger distances. Our comparison allows us to
articulate how PDF flows differ from their hydrodynamic counterparts.

This paper is organized as follows: in section~\ref{model} we
describe the methodology and numerical code that we use for this
study as well as our implementation of the gas, the velocity and
the magnetic field. The results of our simulations are presented
in section~\ref{results}, where we follow the evolution, structure
and stability of our model jets. In section~\ref{discu} we discuss
the implications of our simulations and how they compare with specific
laboratory experiments and astronomical observations. 
Finally, we conclude in section~\ref{conclu}.

\section{Model and  Initial Set-up}
\label{model}

We model PFD and HD jets numerically by solving the equations of 
ideal (i.e. no explicit microphysical diffusivities)
magnetohydrodynamics (MHD) in three-dimensions. In non-dimensional 
conservative form these are given by

\begin{eqnarray}
   \frac{\partial\rho}{\partial t} + \nabla\cdot(\rho\mbox{\bf V}) &= 
   &\dot{\rho}_\mathrm{inj}
   \\
   \frac{\partial (\rho {\bf V})}{\partial t} + 
   %
      \nabla\cdot \left( \rho {\bf V V } + p \, {\bf \hat{I}} + (B^2/2) {\bf \hat{I}} - 
      {\bf B B}\right)  &= &\dot{\bf P}_\mathrm{inj}
	%
   \\
   \frac{\partial E}{\partial t} + \nabla\cdot\left[\left(E
   +p + B^2/2 \right)
   {\bf V}-{\bf B}({\bf V}\cdot{\bf B})\right] &= 
   &\dot{E}_\mathrm{inj} - \Delta(T)
   \\
   \frac{\partial {\bf B}}{\partial t} - \nabla\times( {\bf V}
   \times {\bf B}) &= &\dot{{\bf B}}_\mathrm{inj},
   \label{mhd}
\end{eqnarray}
\noindent where $\rho$, $p$, {\bf V, B} and {\bf \^I} are the gas density,
thermal pressure, flow velocity, magnetic field and
the unitary tensor, respectively.
In (3), \hbox{$E=p/(\gamma-1)+\rho V^2/2+B^2/2$} and represents the
total energy density whereas $\gamma$ is the ratio of specific
heats. We have implemented source terms in the right hand side
of equations~(1)--(4) to account for the injection of mass, momentum, 
total energy, and magnetic flux.  
Since the cross sectional area of the jet base is fixed, these injections are respectively accomplished by
injecting  a mass density per unit time  $\dot{\rho}_\mathrm{inj}$, a momentum flux, 
$\dot{\bf P}_\mathrm{inj}$, total energy flux minus radiation loss
$\dot{E}_\mathrm{inj}-\Delta(T)$ and magnetic field per unit time $\dot{{\bf
B}}_\mathrm{inj}$. 
%
%

We solve these equations using the adaptive mesh refinement (AMR)
numerical code \textit{AstroBEAR2.0}\footnote{
https://clover.pas.rochester.edu/trac/astrobear/wiki} which uses
the single step, second-order accurate, shock capturing 
CTU+CT \citep{CT} scheme \citep{astrobear,bear2}. While
AstroBEAR2.0 is able to compute several microphysical processes,
such as gas self-gravity and heat conduction, we do not consider
these in the present study.

Our  computational domain is defined within $|x|,|y|\leq\,$160\,AU
and 0$\le z \le$400\,AU, where 20\,AU is equivalent to one
computational length unit.  We use a coarse grid of
64$\times$64$\times$80\,cells plus two levels of AMR refinement
which attain an effective resolution of 1.25\,AU. Outflow boundary
conditions were set at the left and right domain faces of both $x$
and $y$, as well as in the upper $z$ face.  At the lower~$z$ face
we combine two boundary conditions: reflective, in those cells
located at $\sqrt{x^2+y^2} \geq r_e$, and magnetic/jet source term
values in those cells located at smaller radii. $r_e=\,$ 31.4\,AU. 
The latter represents the characteristic radius of the energy injection
region, equal to the jets' radius, which is resolved by 24\,cells.

We use BlueGene$/$P\footnote{
https://www.rochester.edu/its/web/wiki/crc/index.php/Blue\_Gene/P}, an
IBM massively parallel processing supercomputer of the Center for
Integrated Research Computing of the University of Rochester, to run simulations
for about 1\,day using 512 processors.

\subsection{Initial conditions} 
\label{ini}

We initialize our simulations with a static gas which has a uniform density of 
200~particles per~cm$^{-3}$ and a temperature of 10000\,K.
Gas is modelled with an ideal gas equation of state and a ratio of
specific heats of $\gamma=\,$5$/$3.
Magnetic fields
are placed in a central cylinder 
of equal radius and height $r_e$. In cylindrical coordinates
the magnetic vector potential is given by
\begin{equation}
{\bf A}(r,z) = \left\{
   \begin{array}{c l}
          \frac{r}{4} [cos(2\,r) + 1][ cos(2\,z) + 1 ] \hat{\phi} + 
          \frac{\alpha}{8} [cos(2\,r) + 1][ cos(2\,z) + 1 ] \hat{k},
             & \mbox{for}~r,z < r_e; \\
          0, & \mbox{for}~r,z \ge r_e,
   \end{array} \right.
\label{apot} 
\end{equation}
\noindent where the parameter~$\alpha$ has units of length and
determines the ratio of toroidal to poloidal magnetic fluxes.  We
use $\alpha=\,$40 (computational length units) which is an arbitrary
choice, yet consistent with the  wound helical magnetic
configurations  expected from accretion discs
\citep[e.g.][]{blandford82,bell96,li06} and produced in high energy
density laboratory experiments of magnetic towers \citep{lab1,lab2}.
Our choice of ${\bf A}$ is 
in part motivated by the work of \citet{li06}. However, in our model, ${\bf A}$
is strictly localized to the central part of the grid.
%

We obtain the initial magnetic field, ${\bf B}^{\mathrm{init}}$,
by taking the curl of ${\bf A}$:
\begin{equation}
   \begin{array}{c l l}
   B^{\mathrm{init}}_r      &=& - \frac{\partial}{\partial z} (A_{\phi})
            = 2 r cos^2(r) cos(z) sin(z), \\
   B^{\mathrm{init}}_{\phi} &=& - \frac{\partial}{\partial r} (A_z)
             = \alpha cos^2(z) cos(r) sin(r), \\
   B^{\mathrm{init}}_z      &=& \frac{1}{r} \frac{\partial}{\partial r} (r A_{\phi}) 
            = 2 cos^2(z) [ cos^2(r) - r cos(r) sin(r) ].
   \end{array}
\label{bfields} 
\end{equation}
\noindent The magnetic field is normalized so that the initial 
  thermal to magnetic pressure ratio $\beta$ 
 is less than unity for $r<r_e$, and unity at $r=r_e$.
In Figure~1 we show profiles of the  initial magnetic field components (top
and middle rows) and $\beta$ (bottom panel) as a function the
distance form the origin.

\begin{figure}
\centering
  \includegraphics[width=.480\textwidth,bb= 60  105 395 290,clip=]{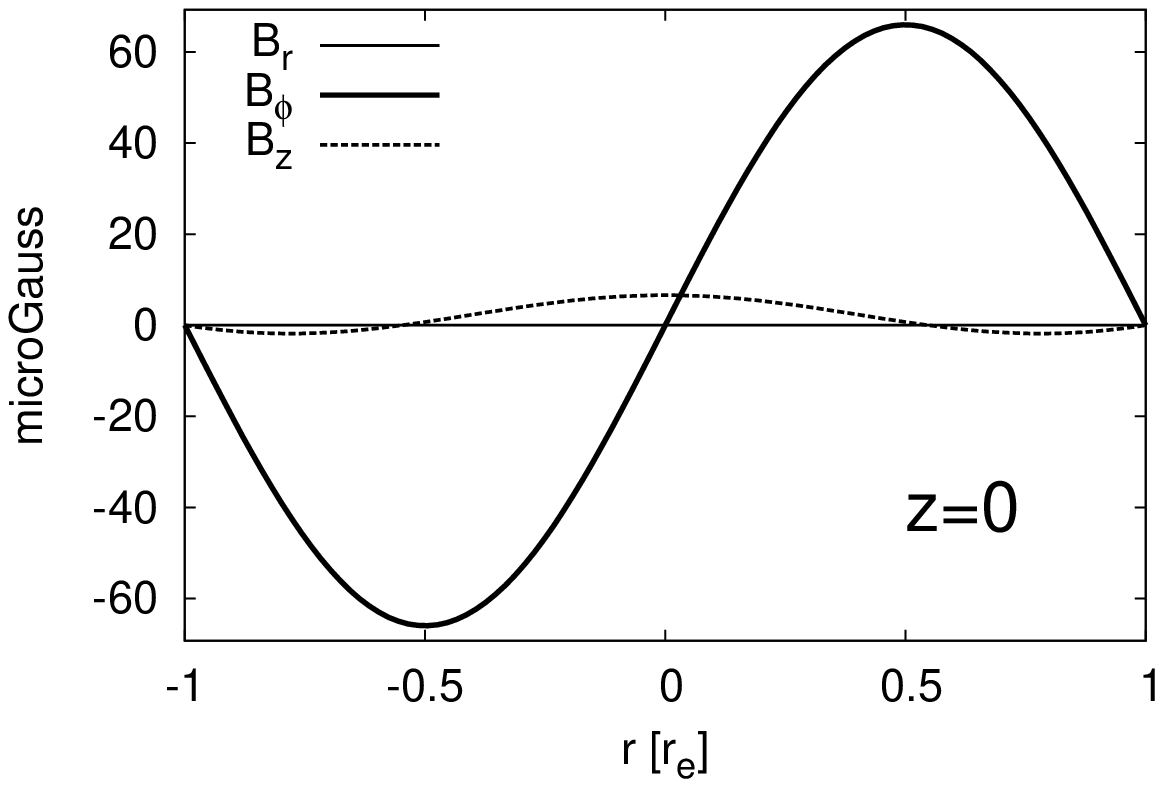} 
  \includegraphics[width=.412\textwidth,bb= 110 105 395 290,clip=]{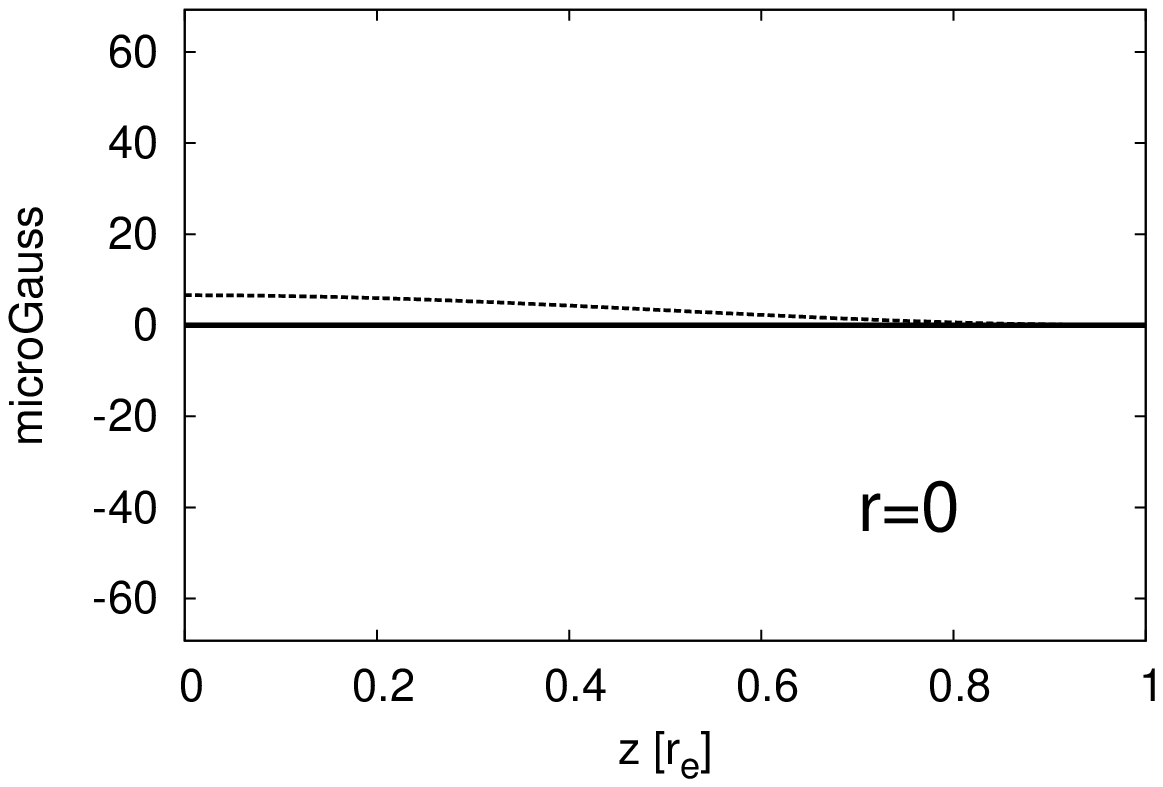} \\ 
  \includegraphics[width=.480\textwidth,bb= 60  65  395 290,clip=]{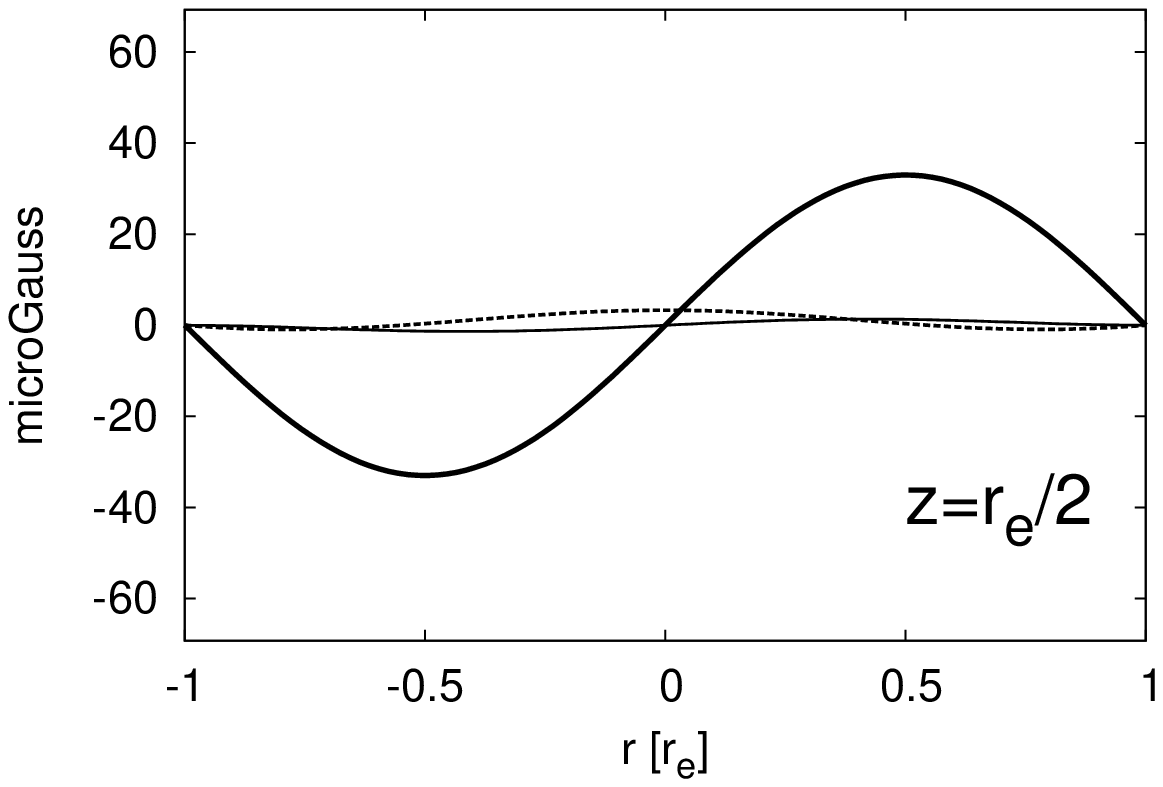} 
  \includegraphics[width=.412\textwidth,bb= 110 65  395 290,clip=]{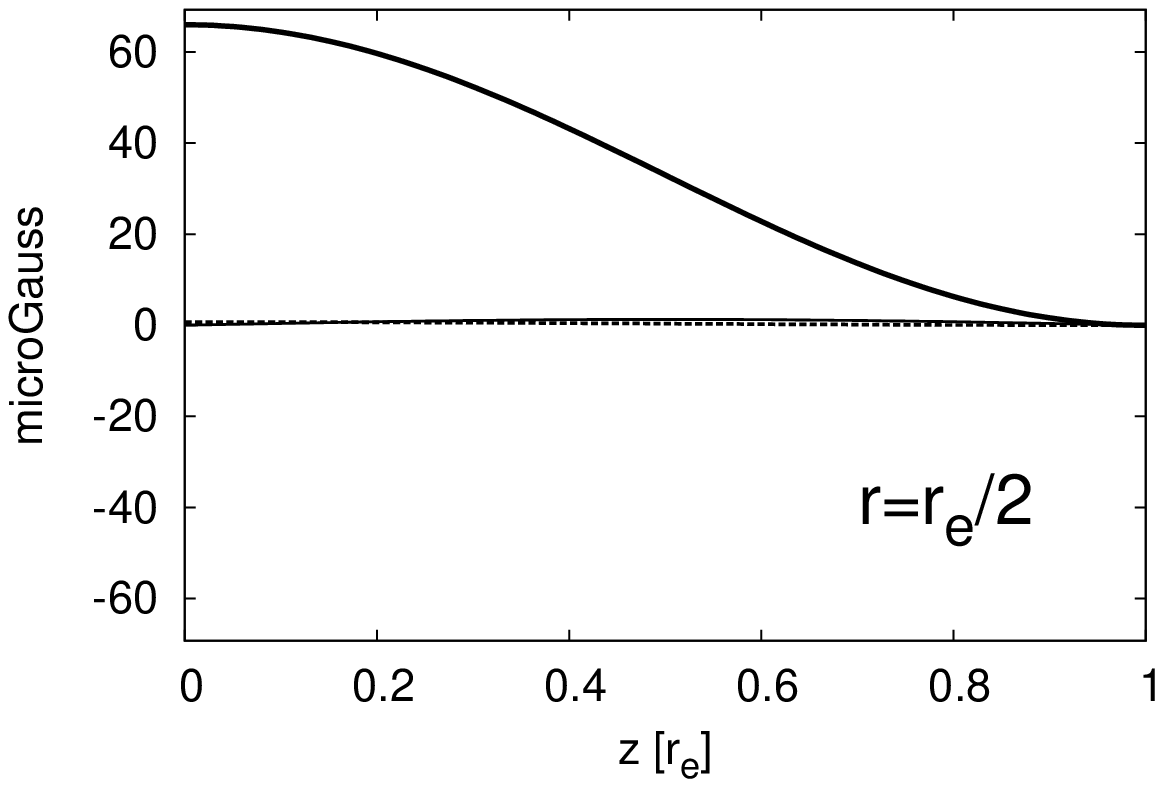}\\ 
  \textit{(a) Field components as a function of $r$ (left column) and $z$ (right column).} \\
  \vskip.3cm
  \includegraphics[width=.4\textwidth]{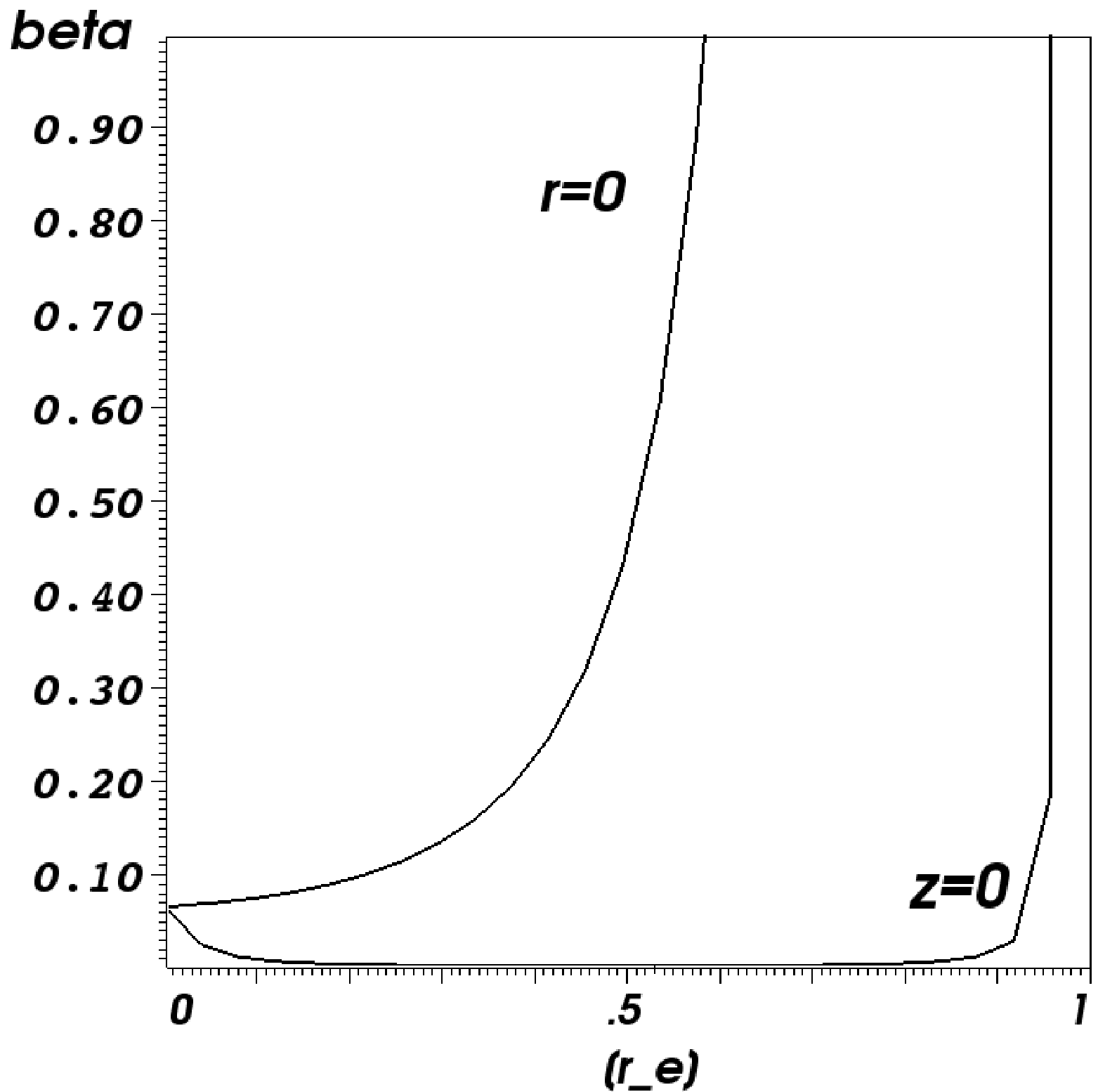}\\
\textit{(b) Plasma $\beta$ (thermal to magnetic pressure ratio).}
  \caption{Magnetic field initial conditions. The field has a
  helical structure which is dominated by the toroidal
  component.
  }
  \label{profiles}
\end{figure}

\subsection{Energy injection} 
\label{inj}

We model jets by continually injecting energy into the
central region of the grid, where $r,z \le r_e$.
Because one of the key goals of this work is to compare 
the observable propagation 
signatures of PFD jets versus HD ones, (e.g. their length, velocity, 
density distribution), we inject either pure magnetic 
energy for the PFD case, hence the name magnetic towers,
or pure kinetic energy for the HD case. 
We will now give details about
the implementation of the jets.

\subsubsection{Magnetic towers}
\label{221}

For these simulations we inject magnetic flux
by adding the initial magnetic
field configuration (\ref{bfields}) to the instantaneous central
magnetic fields, ${\bf B}^n$. i.e.

\begin{equation}
{\bf B}^{n+1}={\bf B}^n+\dot{\bf B}_{\mathrm{inj}} \,dt, \\
\label{ener}
\end{equation}
\noindent where ${\bf B}^{n+1}$ represents the central magnetic
fields ($r,z \le r_e$) corresponding to the next computational
timestep, $dt$ is the current timestep, $\dot{\bf B}_{\mathrm{inj}}={\bf
B}^{\mathrm{init}} \, B_c$ and $B_c$ is the magnetic 
flux injection rate (see below).

For numerical stabilization we also
continually add static gas to the grid within $r,z<r_e$. 
This is accomplished using the expression
\begin{equation}
\rho^{n+1}(r,z)=\rho^n(r,z)+ \rho_c \, |{\bf B}(r,z)|^2 \, dt, \\
\label{mass}
\end{equation}
\noindent where 
$\rho^{n+1}(r,z)$ and $\rho^n(r,z)$ represent the gas densities
corresponding to the next and the current timesteps, respectively.
We set the constant factor $\rho_c$ (which has units of kg\,m$^{-3}$\,s$^{-1}$\,T$^{-2}$)
to be 0.01 computational units.
Hence the average amount of injected gas is of order 
0.001\,$\rho_{\mathrm{amb}}^0$ per unit time,
where $\rho_{\mathrm{amb}}^0$ is the initial 
($t=\,$0\,yr) grid
density of 200~particles per~cm$^3$; very dilute. 
Because of the factor $|B^2|$, 
the distribution of gas provided by (\ref{mass}) matches 
the gradients of magnetic pressure, thus we inject 
more gas at regions where the jet and magnetic cavity densities 
tend to be lower (section~\ref{struc}).

\subsubsection{Hydrodynamical jets} 
\label{hydroinj}

For these simulations we continuously inject
kinetic energy and gas to the cells located at $r<r_e$ and $z<$0,
i.e. within the bottom $z$ boundary. 
This region is equivalent to the base of the magnetic towers (discussed above).
We impose constant boundary conditions in this region, based on the following three assumptions.
(1) The collimation of the HD jet is presumed to have occurred at sub-resolution 
scales. (2) the HD jet is taken to have the same time averaged, maximum propagation speed as the 
PFD magnetic tower, that is
\begin{equation}
v_j = v_z \approx |{\bf B_{\mathrm{max}}}| (4 \, \pi \, \rho_{\mathrm{amb}}^0)^{-1/2}.
\label{assume1}
\end{equation}
\noindent (3) The injected energy fluxes of the HD and PFD magnetic tower jets are taken to be equal,
i.e.
%
\begin{equation}
0.5 \rho_j v_z^3 \, a = ( |{\bf B}|^2 / 8 \pi) \, 
  ( |{\bf B}| ( 4 \pi \rho_{\mathrm{amb}}^0)^{-1/2} ) \, a,
\label{assume2}
\end{equation}
\noindent where $\rho_j$ is the jet's density and $a$ ($=\pi r_e^2$) is the area of the 
energy injection region. Hence, 
\begin{equation}
      \rho_j = |{\bf B}_{\mathrm{max}}|^2  ( 4 \pi v_j^2 )^{-1}.
\label{jet}
\end{equation}
\noindent 
To ensure the condition (\ref{assume2}) at all times, we set 
$B_c=\,$10$/$(1~time computational unit) in equation~(\ref{ener}).
We note that for the HD run, ${\bf B}=\,$0 everywhere and at all
times, and that the values of ${\bf B_{\mathrm{max}}}$ and ${\bf B}$ in
equations~(\ref{assume1})--(\ref{jet}) are taken from the magnetic
tower simulation (above).

\subsection{Simulations} 
\label{models}

We carry out six simulations: three magnetic tower runs
and their corresponding hydrodynamical versions. 

{\bf The adiabatic tower.} This is a magnetic tower model
which we have implemented as described in section~\ref{221}.

{\bf The cooling tower.} This is a magnetic tower model
which is identically to the adiabatic tower except for the addition of 
optically thin cooling  which we have implemented using the tables of 
Dalgarno \& McCray (1972) via the source term $\Delta(T)$ in equation~(3).

{\bf The rotating tower.} This is a magnetic tower model
which is identically to the adiabatic tower 
except for the addition of a rotation profile at the jet base.
This is accomplished by continually driving an azimuthal velocity to the central gas and 
frozen in magnetic fields of the tower. We use a velocity equal to the Keplerian
speed corresponding to a two solar mass star. Specifically we impose
\begin{equation}
   v_{\phi} = \left\{
   \begin{array}{c l}
	\sqrt{G 2 M_{\odot}/r}, &\mbox{for}~r,z < r_e; \\
	                              0, & \mbox{for}~r,z \ge r_e.
	\end{array} \right.
	\label{vel} 
\end{equation} 
\noindent Our choice of two solar masses is arbitrary but within
the expected values for protostellar and young stellar object (YSO)
jet engines \citep[e.g. see][and references therein]{konigl00}.  We
note that the gas in our simulations is unaffected by  gravitational
forces, hence the centrifugal expansion produced by (\ref{vel}) is
only balanced by magnetic pressure gradients. We do not expect
significant dynamical differences with respect to a case in which
gas was affected by gravity because we simulate jets far from the
central star \citep{meier97}. Also, in our magnetic towers the magnetic
fields are quite strong and the magnetic cavities contain very light
gas (see below).

{\bf The HD jet.} This is an adiabatic  hydrodynamical jet model
which we have implemented as described in section~\ref{hydroinj}.

{\bf The cooling HD jet.} This is a hydrodynamical jet model
which is identical to the HD jet except for the addition of the same thermal cooling source
term that we use for the cooling tower run (above). 

{\bf The rotating HD jet.} This is an adiabatic hydrodynamical jet model
which is identical to
the HD jet except for the addition of the base ($r<r_e$ and $z$
within the bottom boundary of the computational domain) rotation
profile described by equation~(\ref{vel}). 

The structure and evolution of the HD, the cooling~HD and the rotating~HD
jet simulations are similar in terms of their global propagation characteristics 
(see section~\ref{hdsec}). Hence, without loss of generality, in what follows
we will only discuss about the adiabatic HD jet.

\section{Results}
\label{results}

\subsection{Plasma structure and evolution}
\label{struc}

\begin{figure}
\begin{center}
$\log(n)$ [cm$^{-3}$]\\
\includegraphics[width=\columnwidth]{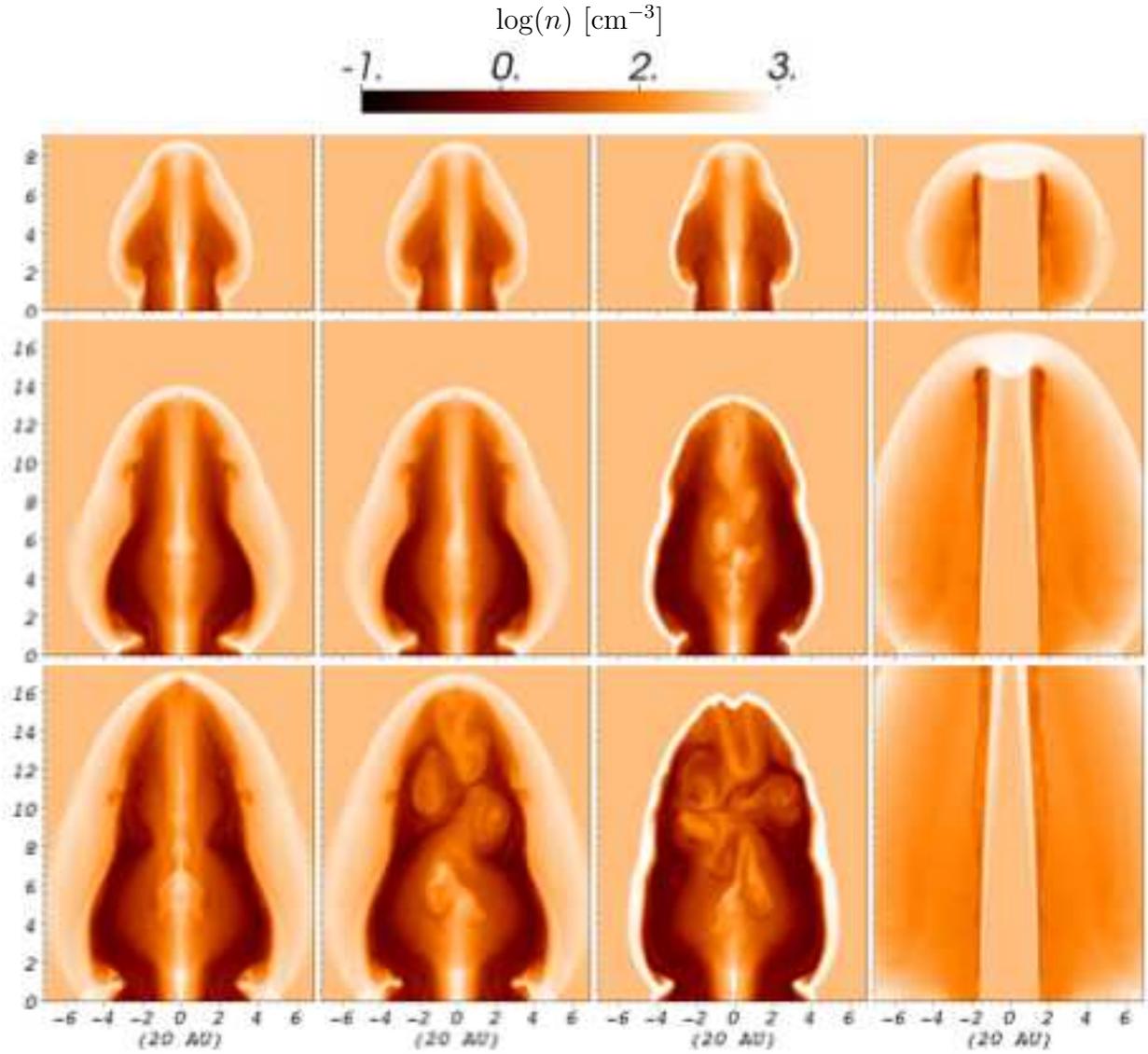}\\
 \caption{Evolution of the plasma gas density. These false color
 logarithmic maps show the magnetic tower structures in the adiabatic
 (1st column), the rotating (2nd column) and the cooling (3rd column)
 cases, and the HD jet structure (4th column). From top to bottom
 time is equal to 42,~84~and 118\,yr.
}
   \label{dens}
\end{center}
\end{figure}

In Figure~\ref{dens} we show the evolution of the plasma with
logarithmic false color particle density maps. From left to right, columns
in the Figure show the adiabatic, the rotating and the cooling
magnetic towers, and then the HD jet. Time increases
downward by row. 
We denote the structures in the simulation as follows, based on the left-most
panel of row~2 (Figure~\ref{dens}):
the jet core (white plasma within $r \la 0.4$); 
the jet beam (lightest-orange plasma within $r \la 1.6$); 
the magnetic cavity (dark-orange plasma within $1.6 \la r \la 4$,
outside the jet); 
the contact discontinuity (CD, thin surface between the magnetic cavity 
and the swept up external medium). Beyond the CD we see 
the (light-orange) shocked ambient plasma.

The simulations show that the initial helically wound magnetic field launches 
PFD jets via magnetic pressure gradients:
The low $\beta$, low density cavities 
expand via the z-gradient of the toroidal magnetic pressure between the tower and ambient medium.
Inside the cavity,  a central jet beam forms collimated by hoop stresses of the toroidal
field (section~\ref{fields}).  The field in the cavity is in turn, radially collimated by the 
pressure of the external high $\beta$ plasma.
 The jets and their corresponding magnetic cavities
expand and accelerate, especially along the $z$-axis. This drives
bow shocks on the external unmagnetized media.   This magnetic tower
evolution is consistent with the analytical model of \citet{bell96,bell03},
as well as with previous simulations of PFD jets and magnetic towers
(see e.g. \citealp{shibata86,nakamura04,li06}).

Comparison of PFD magnetic towers with the HD jet reveals the following 
characteristics (Figure~\ref{dens}).  The towers propagate with
very similar vertical velocities but decelerate, by about 20\%,
relative to the HD jet.   This results because although the towers and
the hydro jet have the same injected energy flux, the towers
produce not only axial but radial expansion. The pre-collimated
HD jet can only expand radially via a much lower thermal pressure. Thus all
of the energy flux in the hydro-case for our set up is more efficiently
directed to axial mechanical power. Moreover, the towers and the hydro
jet show different structures: towers have a thin central jet which is susceptible to instabilities,
whereas the HD jet's beam is thicker,
smoother and stable.  We consistently see lower densities in the
PFD tower cavities than in that of the HD case. 
The laboratory experiment magnetic towers
of \citet{lab1} and \citet{suzuki} also show a magnetic cavity mostly
void of plasma. The gas
distribution inside the cavities shows more complex and 
smaller scale structures in the magnetic tower cases
than in the HD one.

We see that the magnetic towers are affected
by either cooling or rotation after their early expansion
phase. Instabilities develop in their jet beam after~$\sim\,$70\,yr
(section~\ref{sta}). The cocoon geometry of the
cooling case (third column from left to right in Figure~\ref{dens}) is the fattest.  We find that the volume 
of the ambient region which is affected by the towers is smaller in the
cooling case, as expected \citep[e.g.  see][]{frank98,huarte11}.
The above findings imply very different emission distributions for PFD 
and HD dominated jets.  Future studies should address the creation of synthetic observations to assess these differences.

\begin{figure}
\begin{center}
\includegraphics[width=\columnwidth]{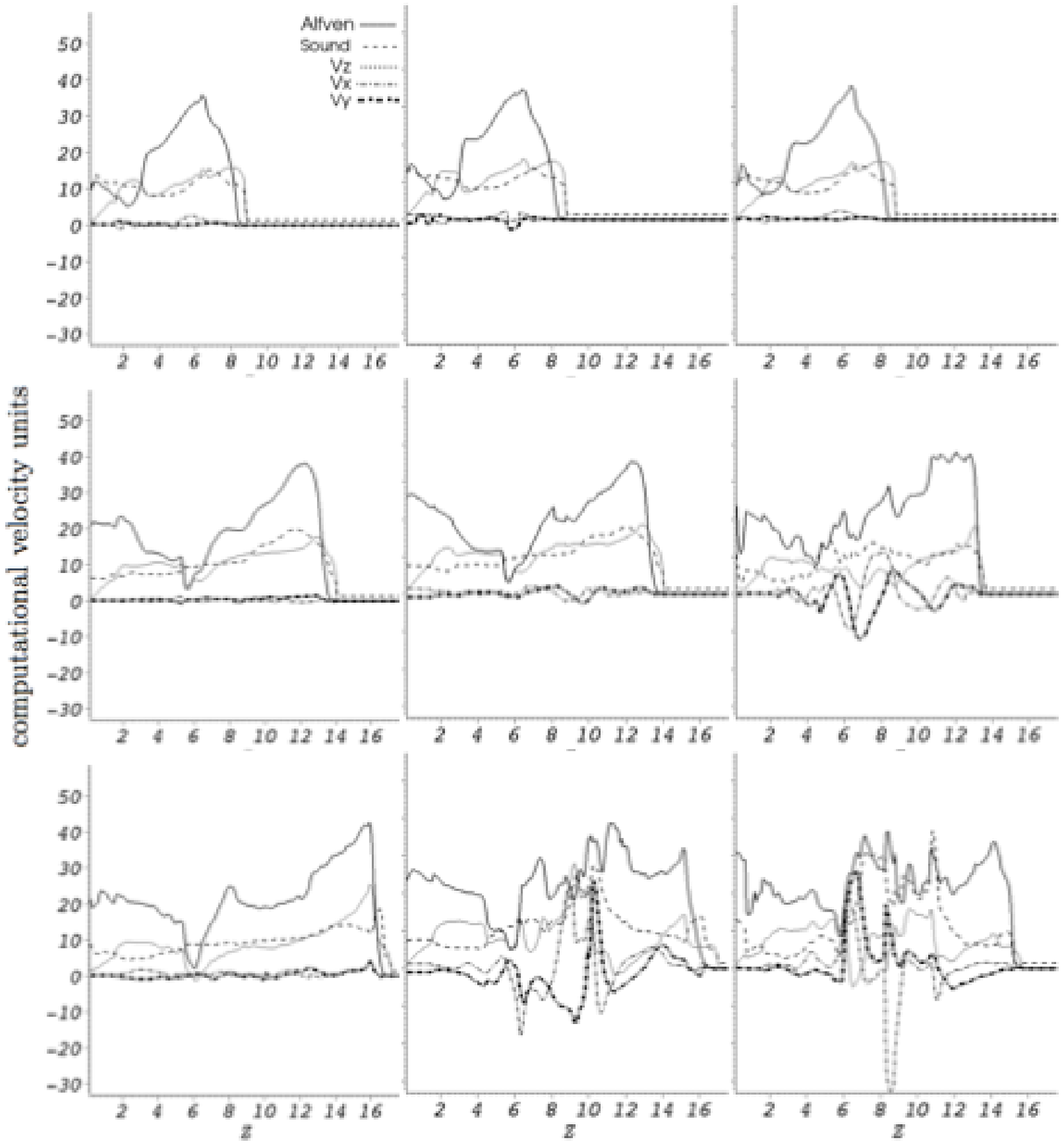}\\
 \caption{Evolution of the plasma velocities along the jets' axis. These are \capta.
 Each computational velocity unit is equivalent to 9.1\,km\,s$^{-1}$. 
}
   \label{vels}
\end{center}
\end{figure}

The evolution of the magnetic towers' gas density is consistent
with that of their compressive MHD  and hydrodynamic waves and
shocks. In Figure~\ref{vels} we show profiles of the relevant
velocities of the towers ($v_x, v_y, v_z$, the sound speed and the
Alfv\'en speed) along the jet axis, $r=\,0$, as a function of
cooling, base rotation, and time.  During their early stable propagation phase,
the jet cores
are mostly sub-Alfv\'enic and trans-sonic,
independent of cooling or rotation.  Fast-forward compressive MHD
(FF) and hydrodynamic bow shocks are evident 
in the ambient medium, ahead of the jet heads.

Some evolutionary features are worth noting.  From figure 3 we see
the FF shocks steepen in time (compare top to middle and middle to
bottom rows). The swept of shells of unmagnetized ambient medium
become relatively thin when radiative cooling is included (right
column: compare top to middle and middle to bottom rows).  The
adiabatic and rotating cases show regions within the lower half of
the jet where the flow speed is super-Alfv\'enic. Such regions are
bounded by the reverse and the forward slow-modes of compressive
MHD waves, and characterized by high thermal to magnetic pressure
ratios (section~\ref{stable2}).

At $t \ga \,$90\,yr the distribution of waves and shocks of both
the rotating and cooling cases (bottom row, middle and right
columns) is significantly affected. This is due to the
growth of non-linear current-driven instabilities (section~\ref{sta}).
Possibly, pressure driven modes coexist with the current driven ones
in regions of high $\beta$.
We see fast, though mostly sub-Alfv\'enic, azimuthal velocities, 
in the central parts of these jet cores.

To clarify the effect of cooling on our magnetic towers we present
temperature maps in Figure~\ref{temp} below. We model radiation
losses using~(3), where $\Delta T \propto \rho^2 \Lambda(T)$ and
$\Lambda (T)$ is taken from the tables of \citet{dm}. Figures~2
and~4 help to form a complete picture of the cooling strength. In
the non-cooling cases we see shocked ambient medium at temperatures
of $T \sim\,$10$^5$\,K.  This material forms an extended shell surrounding
the magnetic cavity formed by inflowing Poynting flux.  In the cooling
case this shocked shell of ambient gas has cooled significantly to
temperatures of $T \le \,$10$^4$\,K.  The cooling has decreased the pressure
in this region on the shell now becomes both thin and dense. Note
we also see low temperature via cooling occur in the jet beam and
the knots that form once the beam becomes unstable.

\begin{figure}
\begin{center}
$\log(T)$ [K] \\
\includegraphics[width=\columnwidth]{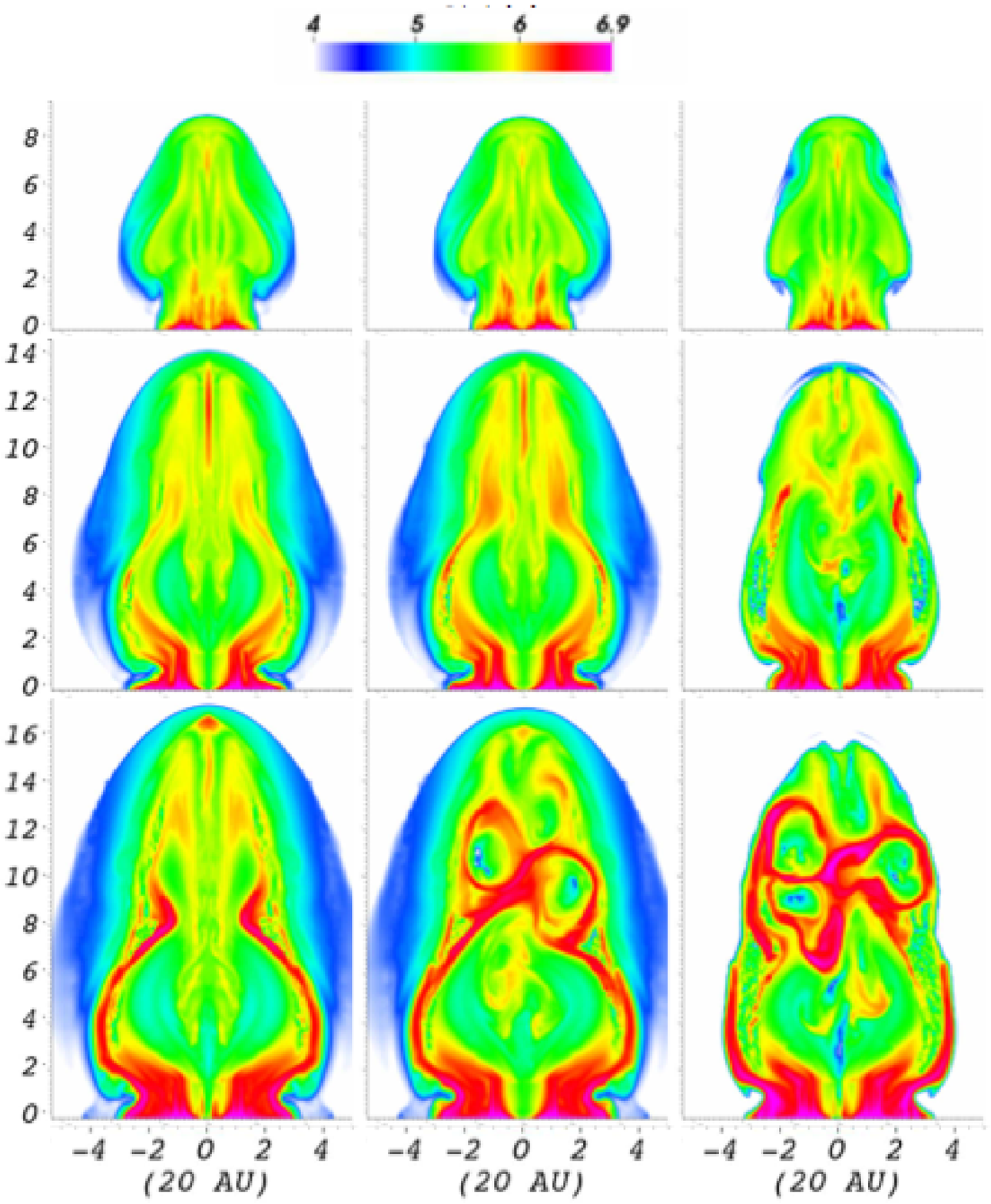}\\
\caption{Evolution of the towers' temperature. These logarithmic color maps show \capta.
}
\label{temp}
\end{center}
\end{figure}

\subsection{Magnetic fields and current density}
\label{fields}

\begin{figure}
\begin{center}
$|B_y|^2 / \sqrt{B_x^2+B_z^2}$ \\
\includegraphics[width=\columnwidth]{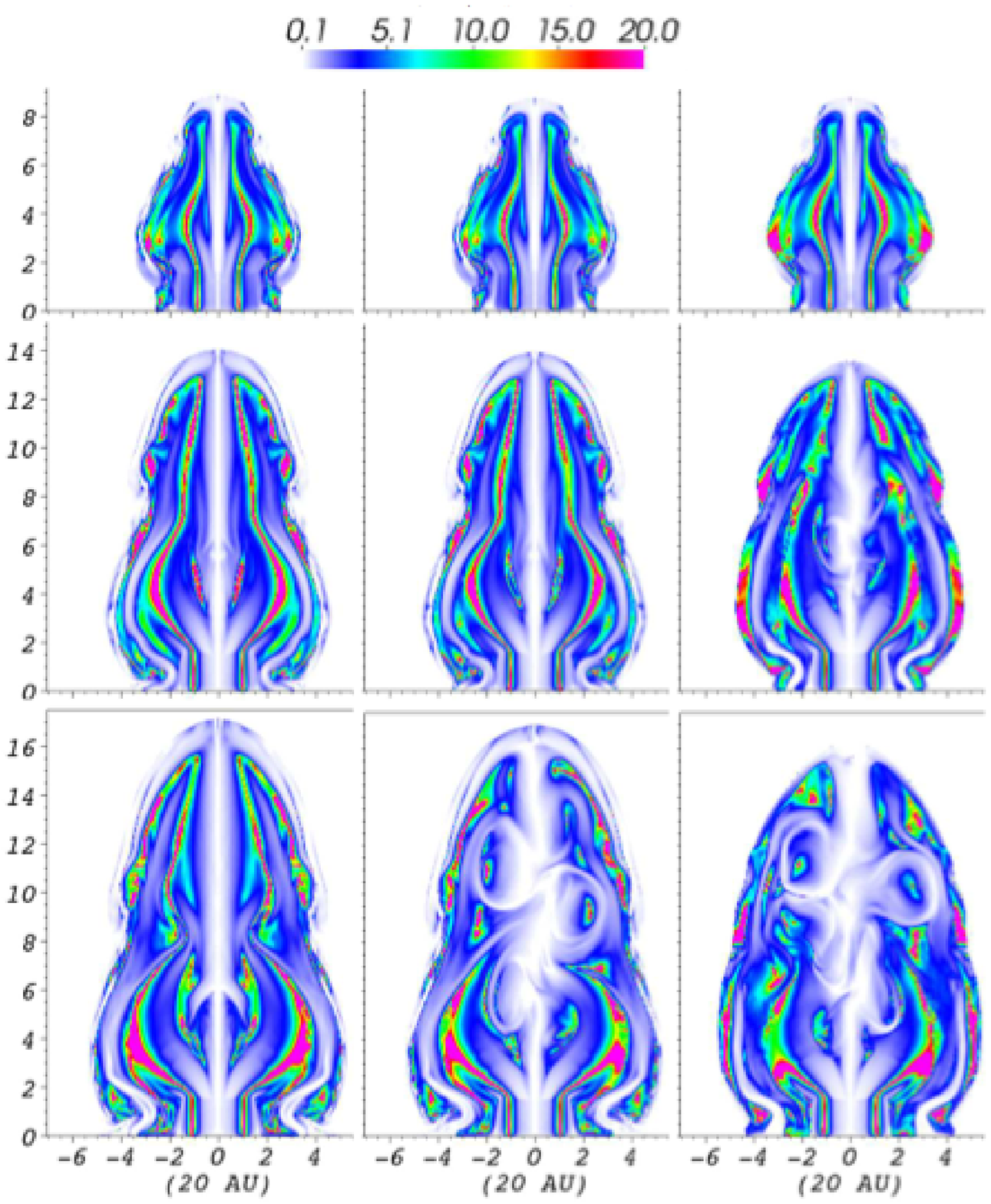}\\
\caption{Evolution of the towers' magnetic fields. This is the ratio of the toroidal component 
over the poloidal one. $B_y = B_{phi}$ and is perpendicular to the maps. 
These linear colour maps show \capta.
}
\label{Bratio}
\end{center}
\end{figure}

In Figure~\ref{Bratio} we show the distribution of the towers'
magnetic fields on the plane that contains the jets' axis. 
These are linear color maps of the absolute value
of the toroidal to poloidal field component ratio.  From
left to right we show the adiabatic, the rotating and the cooling
cases, respectively, and time increases downwards with row.  We
see that the magnetic flux changes sign along the radial direction. In general there
are four main nested surfaces or layers of magnetic field lines
(e.g. see middle row, left column panel): 
at the very core we see predominately poloidal (vertical $B_z$) 
fields surrounded by a surface of primarily toroidal (azimuthal $B_\phi$) flux. These field
components represent the central core of the jet plasma column.  They
are collimated by two outer magnetic surfaces. The smaller  
of these two is dominated by poloidal lines, whereas the larger one is dominated by toroidal lines.
These outer field lines are collimated by the thermal pressure and inertia of the external 
media.

As expected, the geometry of the towers' magnetic fields changes
in time. Initially, the field lines have a highly wound helical
configuration (section~\ref{ini}). The magnetic pressure is very
high and unbalanced in the vertical direction. The toroidal field
lines thus move away from each other and the magnetic towers rise.
The injection of magnetic flux 
sustains a non-force-free configuration at the base of the tower. ``New''
field lines push the ``old'' ones upwards then. The latter stretch and
expand radially, making way for, and collimating, the jets' new
field lines.  

\begin{figure}
\begin{center}
Magnetic field strength [$\mu$G]\\
\includegraphics[width=\columnwidth]{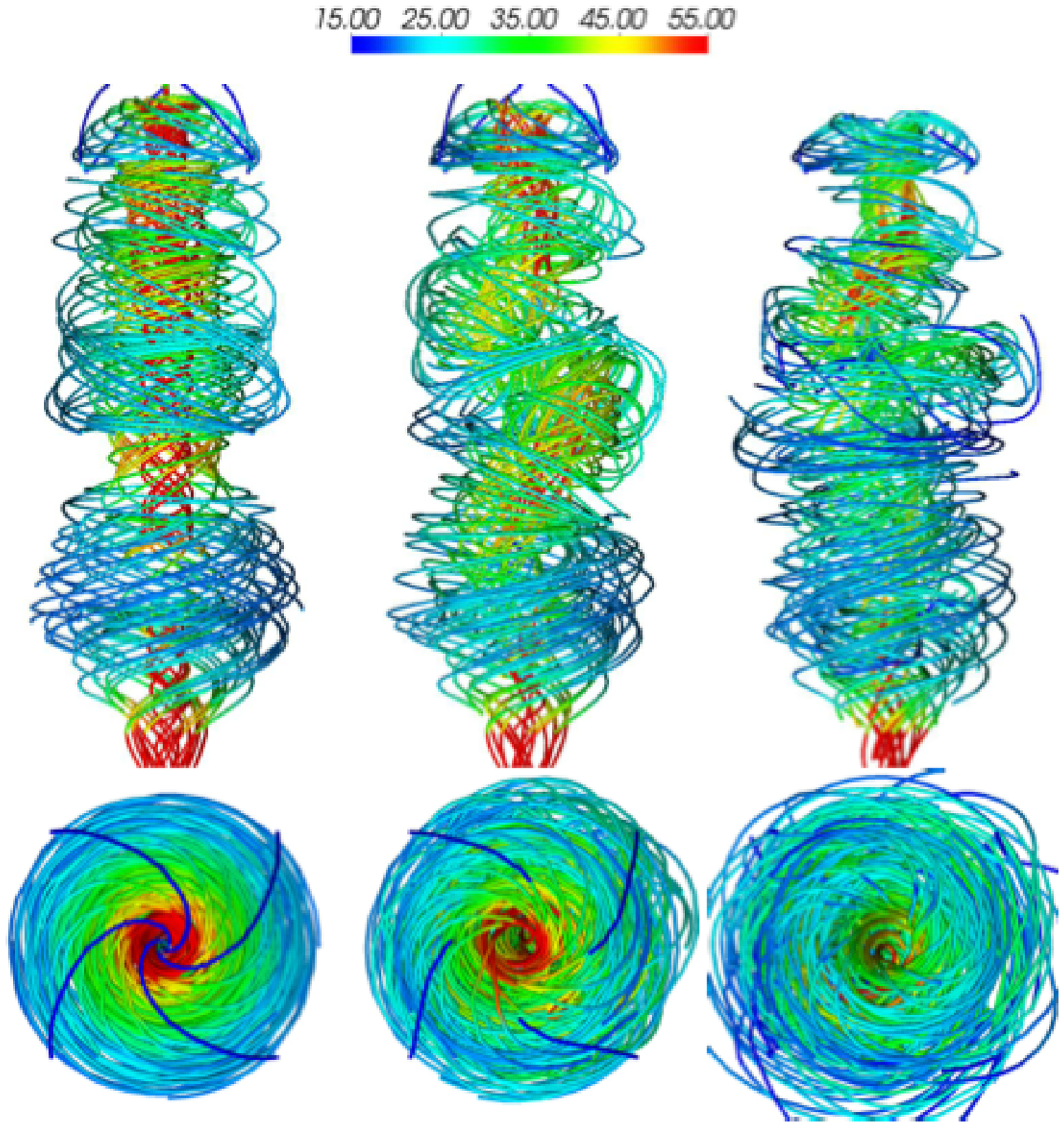}\\
\caption{Central ($r \la \,$ 1.2 $r_e$) magnetic field lines at
$t=\,$118\,yr.  From left to right
these are the adiabatic, the rotating and the cooling magnetic
towers, respectively. Bottom panels show an upper view, pole-on.
Open field lines are a visualization effect.}
  \label{lines}
\end{center}
\end{figure}

After the towers early expansion phase ($t \ga \,$90\,yr),
we find, in agreement with the results of the previous section,
that the jets of both the cooling and the rotating cases are
affected by instabilities (section~\ref{sta}).  The final magnetic
structure of the towers is clearest in the field line maps of 
Figure~\ref{lines}. These are the lines in the central part ($r
\la \,$ 1.2 $r_e$) of the adiabatic (left), the rotating (middle)
and the cooling (right) towers at $t=\,$118\,yr.  The top and
bottom panels shows the towers edge-on and pole-on, respectively.
The adiabatic case shows quite ordered helical field lines and the
strongest jet fields (red color) of all the towers.  We also see
that toroidal field lines tend to pile~up at the tower's tip.
Such a concentration of lines causes acceleration of 
the plasma the tip of the adiabatic jet to supersonic speeds 
(see bottom, left panel in Figure~\ref{vels}, $z\approx\,$15).
In contrast, the cooling tower (right panel) shows the weakest and most disordered
field lines.  
%
%
The middle and right panels show clear differences
between the cooling and the rotating cases. The instabilities that
develop in these towers are clearer in the rotating case (middle
column; setion~\ref{sta}).

\begin{figure}
\begin{center}
$J_z/|max(J_z)|$ \\
\includegraphics[width=.8\columnwidth]{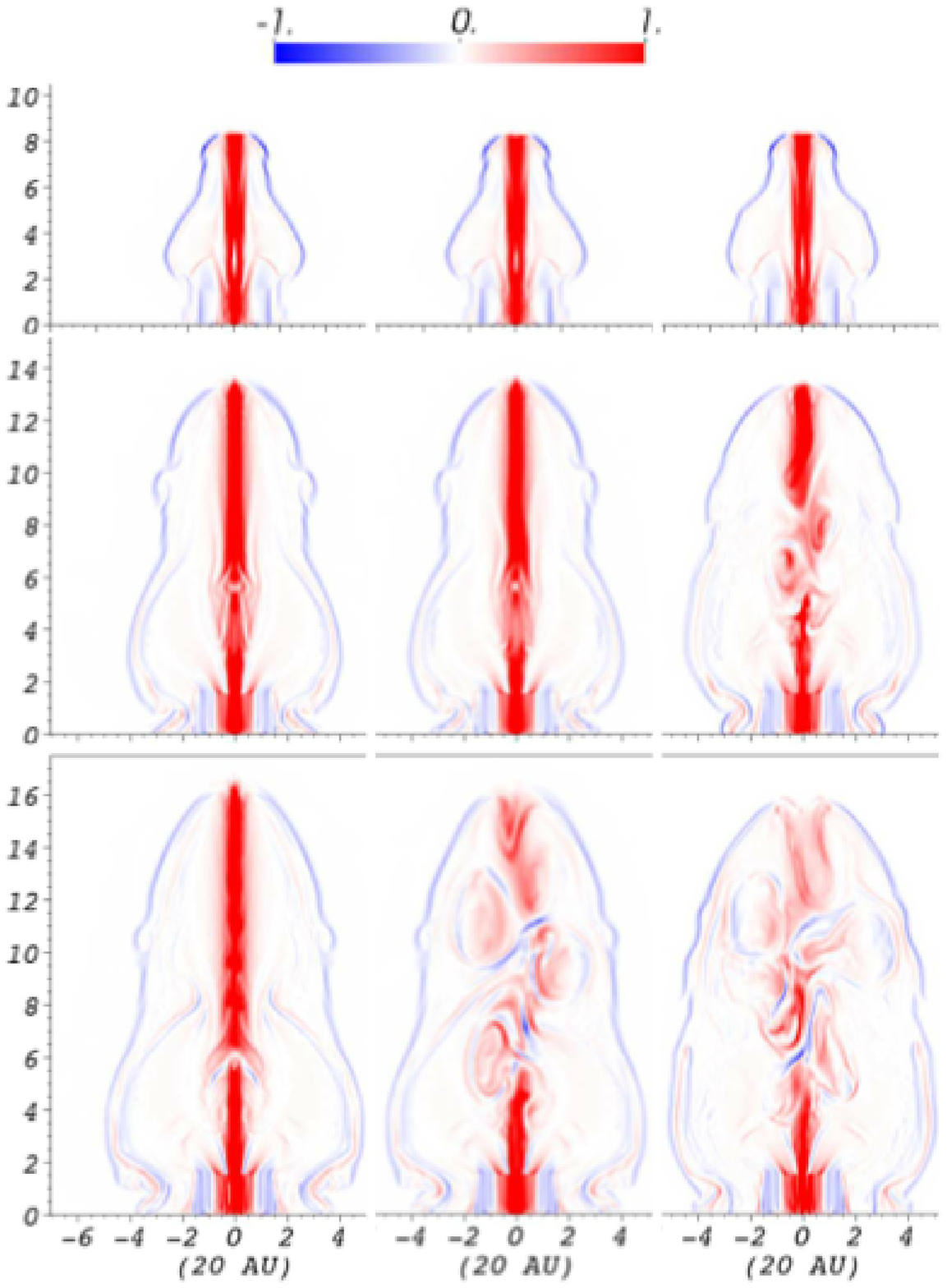}\\
\caption{Evolution of the axial current density. 
$J_z$ has been normalized to the absolute value of its maximum value, $|max(J_z)|$, 
for display purposes. These linear color maps show \capta.
}
\label{curr}
\end{center}
\end{figure}

The magnetic fields are ultimately sustained by  electric
currents. 
In Figure~\ref{curr} we show the evolution of the axial
current density, $J_z$ (panels in this Figure are arranged as in
Figure~\ref{Bratio}).  As expected we see a clear correlation between
the distributions of the axial current density and the magnetic
field. The jets carry a high axial current (red region) which is contained 
within a current-free region (white one) at larger radii. The main part of
the return current (blue region) moves along the contact surface of the towers'
cavity. This forms a closed circulation current system which is
consistent with previous simulations of PFD jets (see e.g.
\citealp{lind89,lery00,nakamura04,lab2}) and the magnetic tower
laboratory experiments of \citet{lab1} and \citet{suzuki}.  We note however that both
the magnetic field and the current density are strictly localized in our model,
i.e. no components of the current in the external medium. This is a characteristic  feature
of magnetic towers.

We see that the current $J_z$ is also affected by the instabilities that develop
in the rotating and cooling towers after their early expansion phase
(bottom row, middle and right columns). The effect of instabilities is most pronounced
in the jet beam. As the jet breaks up into clumps the current becomes more localized.  
Numerical reconnection allows some of the sections of tangled fields to 
become isolated however the overall flow of axial current density 
continues as does the outer sheath of return current.

\subsection{Energy Flux}
\label{stable}

To  study the relative magnetic vs. kinetic  energy content of our magnetic towers 
we compute the Poynting flux, $f_P$, and the kinetic flux, $f_k$, 
defined respectively as
\begin{equation}
   \begin{array}{c l l}
   f_P = & \int\limits_s \, [{\bf B \times (V \times B ) }]_z \, dS, \\
   f_k = & \int\limits_s  \, \frac{1}{2} \, \rho \, |{\bf V}|^2 \, V_z \, dS.
   \end{array}
\label{fluxratio} 
\end{equation}
\noindent The above integrals are taken over the area of the jet beams.

\begin{figure}
\begin{center}
~~~~~~~~~~~$\log \left| Q({\bf x},t) \right| = \log \left| \frac{f_P}{f_{k}} \right| $ \\
\includegraphics[width=.35\columnwidth]{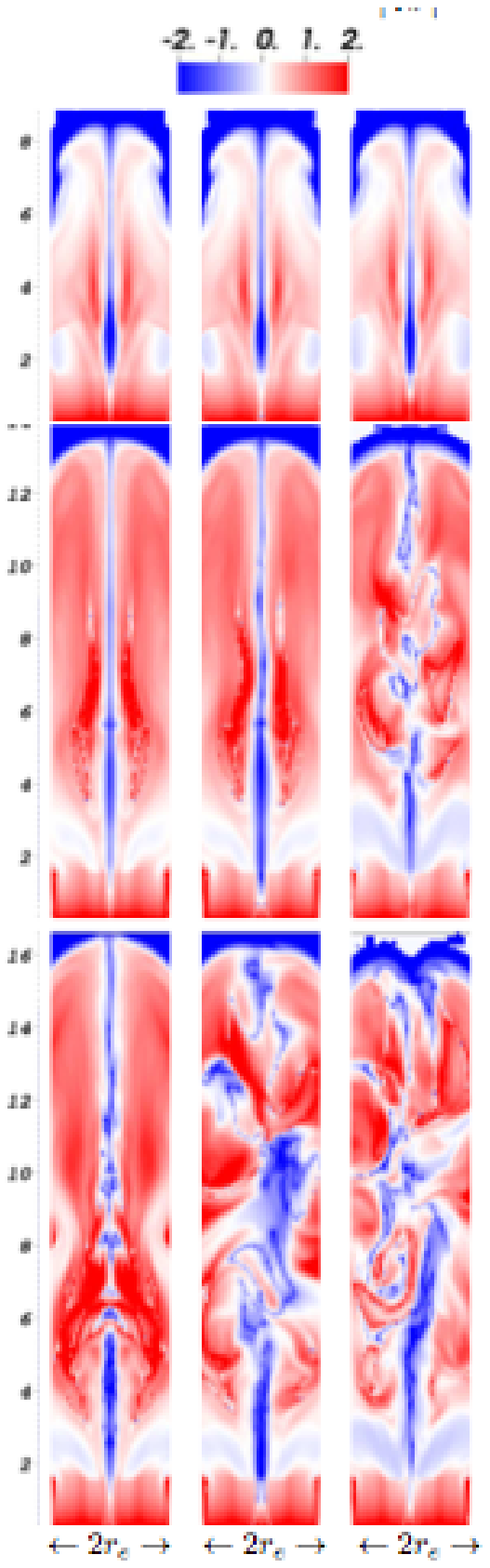}\\
\caption{Distribution and evolution of the jet beam Poynting to kinetic flux ratio. 
These logarithmic maps show the jets of \capta.
}
\label{pfd}
\end{center}
\end{figure}

\begin{figure}
\begin{center}
~~~~~~~~~~ $ \log \left| f_k \right|$ vs $\log \left| f_P \right| $ \\
\includegraphics[width=\columnwidth]{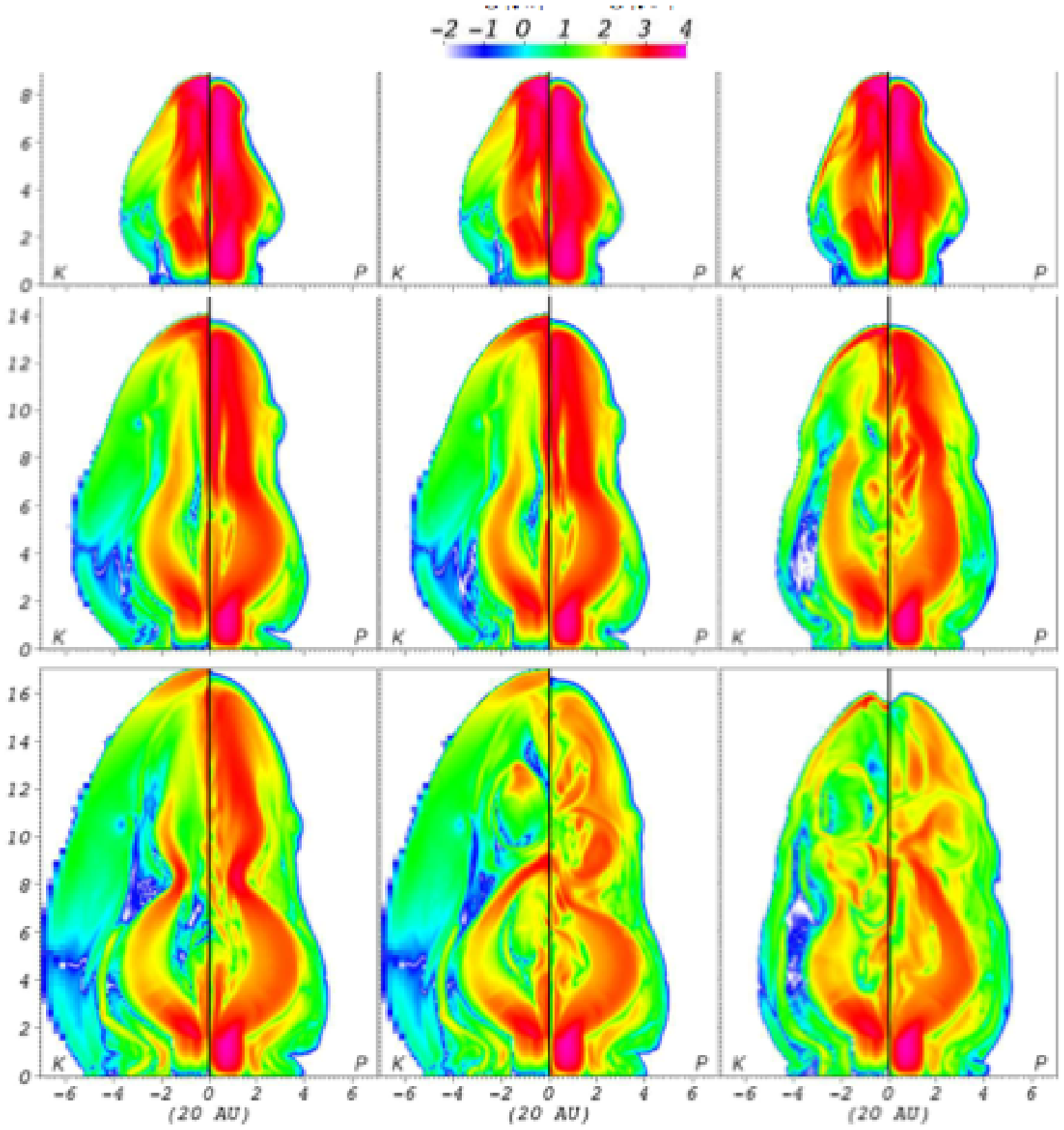}\\
\caption{Distribution and evolution of the towers' kinetic energy
(left) and Poynting (right) polar fluxes in computational units.
These logarithmic maps show \capta.
}
\label{2fluxes}
\end{center}
\end{figure}

In Figure~\ref{pfd} we show logarithmic color maps of the distribution
of the jet Poynting to kinetic flux ratio, $\log \left| Q({\bf x},t) \right|$,
where $Q({\bf x},t)= f_P/f_k$, as a function of colling, jet base rotation and time. 
The maps show that only the core of the jets is dominated by kinetic
energy flux ($Q < \, $1, blue region) while the bulk of the beam is PFD 
($Q > \, $1, red region)
for all the cases (i.e. adiabatic, rotating and cooling).
This distribution is consistent with that found in 
the laboratory jets of \citet[][section~\ref{discu}]{lab1}. 
We confirm that our magnetic towers are indeed PFD. 
We note that the dark red stripes in Figure~\ref{pfd}
correspond to regions where the toroidal field components 
are particularly strong (see Figure~\ref{Bratio}).
%
   To stress and clarify this point we also
   show logarithmic maps of $f_k$ (left panels) and $f_P$ (right panels)
	in Figure~\ref{2fluxes}.
%

We find that the time average mean $Q$ of our magnetic tower beams
--\,averaged over the adiabatic, cooling and rotating cases\,-- 
is $\sim\,$6. This is about 2$/$3 of the time average mean $Q$ in
the magnetic towers simulated by \citep[][see their Fig.~3b,
bottom]{kato}. We note that the spatial distribution of $Q$ in both
our and their simulations is not isotropic and time-dependent. Early in
the evolution of our towers $Q$ is axisymmetric however the growth
of the kink instabilities eventually leads to the development to far more spatial
variability in $Q(x,y,z)$.

Our simulations show the ratio of Poyting flux to kinetic energy
flux is always greater than unity for the magnetic tower ($Q >\,$1).
This should be compared with the models of jets created by
magneto-centrifugal (MCL) processes. While MCL jets begin with $Q>\,$1
on scales less than the Alfv\'en radius, in the asymptotic limit the
kinetic energy flux comes to dominate the flux electromagnetic
energy leading.  Simulations of MCL launching in which the flow is
cold and gas pressure can be ignored show typical values of $Q \sim \,
$0.7 at observationally-resolved distances from the engine (Krasnopolsky
et al. 1999, 2003). We leave a detailed comparison of PDF and MCL jets
for the future.

\subsection{Forces}
\label{stable2}

\begin{figure}
\begin{center}
$\log (\beta)$ \\
\includegraphics[width=.9\columnwidth]{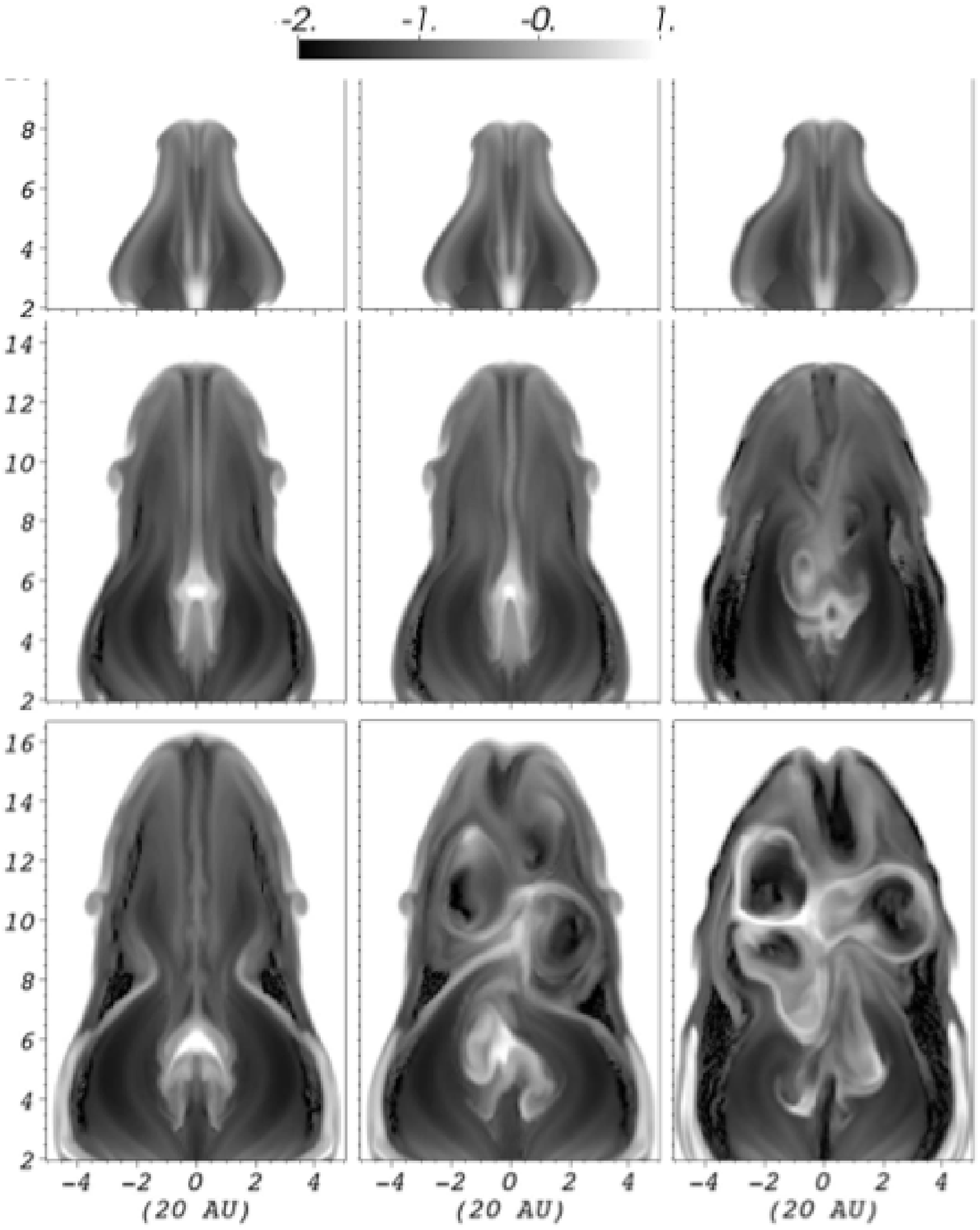}\\
\caption{Evolution of the thermal to magnetic pressure ratio. These logarithmic 
grey scale maps show \capta.
}
\label{beta}
\end{center}
\end{figure}

Magnetic towers expand due to a combination of magnetic, thermal
and inertial forces. In Figure~\ref{beta} we show the thermal to
magnetic pressure ratio, $\beta$, using logarithmic grey scale maps
(arranged as in Figure~\ref{Bratio}). 
We find $\beta$  for the magnetic towers is generally and consistently
well below unity.  

The adiabatic and rotating cases (left and middle columns, respectively)
do show regions where $\beta > \,$1 close to $(r,z)=($0,6). Such
regions are located between the reverse an forward slow-mode
compressive MHD waves (Figure~\ref{vels}, left and middle columns), 
and filled with subsonic,
weakly magnetized plasma.  This high-$\beta$ region is strongly
affected by cooling (right column) which reduces the thermal energy
(see also Figure~\ref{temp}).
Hence the total pressure of the surrounding plasma becomes further
dominated by the magnetic component, and it collapses yielding higher
compression ratios than the adiabatic case. 
The field in the cooling case also takes on a configuration amenable 
to instability.  Thus the plasma in the high-$\beta$ jet-core region plays 
a critical role on the overall stably of PFD outflows.

\begin{figure}
\begin{center}
\includegraphics[width=\columnwidth]{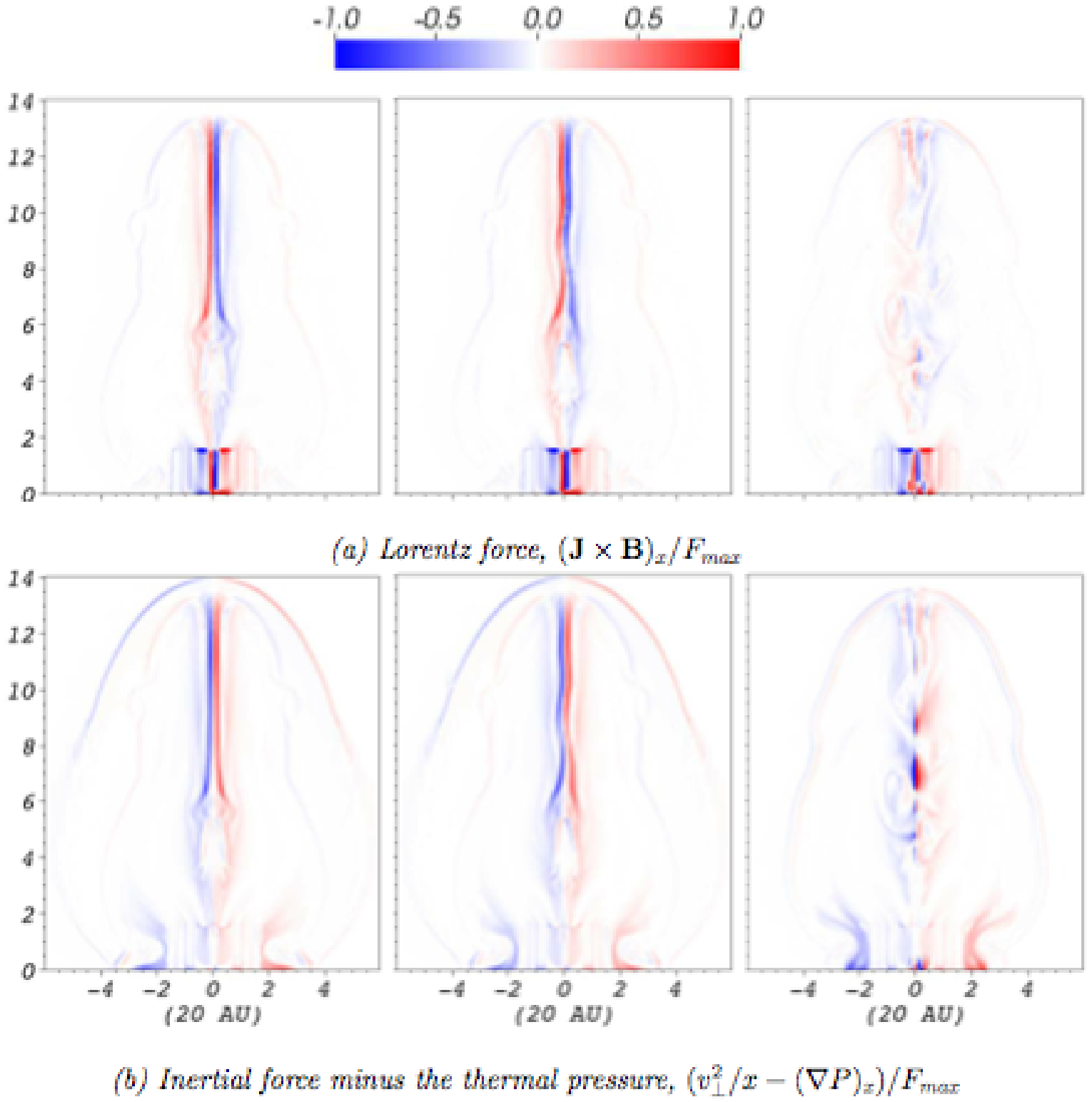}\\
\caption{Radial forces at the intermediate evolutionary phase ($t=\,$84\,yr) of the towers.
Forces are normalized to the maximum value, $F_{max}$. The horizontal axis is $x=r$,
the vertical axis is $z$ and $v_{\perp}$ is perpendicular to the maps.
}
   \label{fforces}
\end{center}
\end{figure}

In Figure~\ref{fforces} we show the radial component of the forces
in the magnetic towers during their intermediate
evolutionary phase. Form small to large radii
these linear color maps show
the jet core (dark colored regions), 
the jet beam edge (light colored regions),
the cavities' central force-free region (white region), 
the CD (light colors), and finally the swept-up ambient gas (outer-most light features, bottom row) of the towers.  
In general these figures show that the inward Lorentz force (top panels) 
is slightly stronger than 
both the inertial (or specific centrifugal $v_{\perp}^2/x$), and 
the thermal pressure, $({\bf \nabla} P)_x$, forces 
which push plasma outward.
%
   This fact is consistent with the results of 
   Takeuchi et al. (2010) and Ohsuga \& Mineshige (2011).
%

Figures~\ref{curr} and~\ref{beta} show the character of the force 
density distribution responsible for confining the jets
and their cavities.  The jets are
self-confined in the current-free region located at a few
jet radii from the core (i.e. hoop stress).  At larger radii, near the towers' contact
surface, which is also the return current surface (blue outer region
in Figure~\ref{curr}), the magnetic pressure is weak and thus it
only requires a mild ambient pressure (light-grey outer region,
Figure~\ref{beta}) to confine the outer part of the towers.

In section~\ref{struc} we saw that the magnetic towers decelerate
with respect to the HD jet.  This can be understood with
the bottom panel in Figure~\ref{fforces} where we see that the
magnetic flux that is injected into the towers (within model $r \la
1.5$; section~\ref{221}) causes not only axial ($z$)
expansion, but also radial expansion via magnetic pressure.
In contrast, the kinetic-energy flux in the HD jet (not shown) 
is overwhelmingly axial.

\subsection{Stability} \label{sta} 

The structure and expansion of our PFD magnetic tower jets are affected
by current-driven instabilities. We see evidence of the pinch,
$m=\,$0, the kink, $m=\,$1, and the $m=\,$2 normal mode perturbations. These
are expected in expanding magnetized plasma columns and consistent
with the models of \citet[][and references therein]{nakamura04} and
Ciardi et~al. (2007) --\,but see also \citet{song}\,-- and also with the
laboratory experiments of \citet{lab1} and \citet{suzuki}.  We find that the kink 
perturbations grow and lead to instabilities in the cooling tower, 
firstly, and later also in the rotating tower.

Perturbations with modes $m=\,$0 and~2 develop in the adiabatic
jet after expanding for $\sim \,$80\,yr
($\sim$70\% of the total simulation running time).
These are caused by radial gradients in the magnetic fields located
within the jet beam, at the boundary of the current-free, force-free region. The
thermal and magnetic components of the total pressure balance
each others' perturbations locally. As a result, the core of the
adiabatic jet becomes a helical column with an
elliptical cross-section.  The growth rate of these m=0 and m=2 perturbations
is $>$120\,yr, which is longer than the simulation
final time. Thus we are seeing only the linear develop of the modes. 
Finally we note resemblances (Figure~\ref{dens}, bottom left panel) 
to structures seen in the S$_{\rm II}$
emission distribution of the jet in HH~34 \citep[][and references
therein]{reipurth02}.

The central part of the towers' jet beams
are high beta plasma
columns where $|B_{\phi} / B_z | \ll 1$
(Figure~\ref{Bratio}, $B_{\phi}=B_y$). 
To understand their development we can appeal to standard Kruskal-Shafranov criterion for the kink instability, namely
\citep{book}

\begin{equation}
   \left| \frac{B_{\phi}}{B_z} \right| > | (\beta_z - 1)k r_{jet}  |,
   \label{insta}
\end{equation}
\noindent where $\beta_z=2 \mu_0 P / B_z^2$ and $k^{-1}$ is the
characteristic wavelength of the current-driven perturbations. In
Figure~\ref{vels} (right column) we see that the cooling jet's core
shows $\beta_z \sim\,$1 ($z \sim \,$4--5 at time$=\,$84\,yr, and 
$z \sim \,$6--11 at time$=\,$118\,yr).  This means that the
cooling tower does not have sufficient thermal energy, in comparison
with the adiabatic and rotating cases, to balance the magnetic
pressure kink perturbations.
This is consistent with what we see in the towers' density and temperature maps,
Figures~2 and ~4.
In addition, we see that the jet core radius of the
cooling tower is about 20\% smaller than that of the adiabatic tower.
This is consistent with what is found in laboratory experiments of
magnetized supersonic jets, in which outflows with different cooling
rates are compared (Ciardi et~al. 2012, in prep). Both of these
effects (thermal energy losses and core radial compression) reduce
the right hand side of (\ref{insta}), making the system more
susceptible to the growth of kink instabilities in the cooling tower.

For the rotating case, we
find that rotation at the base of the jet beam, equation~(\ref{vel}), 
causes a progressive, slow amplification of the
toroidal magnetic field component of the jet.
This process is evident in the
four panels 
at the bottom left 
of Figure~\ref{vels}, where we see that in general the Alfv\'en
speed is higher in the rotating case (middle column) than in the
adiabatic one (left column).  This growth is likely sufficient to
amplify the left hand side of equation~(\ref{insta}) pushing the jet into the
unstable regime. Since the field grows linearly with the differential
rotation,  the growth rate is likely  proportional  to the imposed
amount of rotation. We have not tested this as we have
used only one vale of the differential rotation.

Note that the towers are not completely destroyed even when unstable, and the amplitude of
the kink perturbations in the jet are about twice its  radius  (Figure~\ref{lines}), 
in agreement with the Kruskal-Shafranov
criterion \citep{kruskal,shafranov}.  

\subsection{The HD cooling and HD rotating cases}
\label{hdsec}

In addition to the
four simulations presented above, we have carried out two variations
of the HD jet run: one with cooling, and one which
has a base rotation profile which follows equation~(\ref{vel}),
just as in the rotating magnetic tower run (section~\ref{models}). We found
that the results of the cooling HD jet simulation were
consistent, as expected, with those found in previous similar studies,
i.e. thin jet-produced shocks with high compression factors
\citep[see e.g.][]{frank98,hardee97}. For the regimes we have studied, the propagation
and structure of the HD jets is affected by both cooling and rotation in ways which have
been studied before and which do not alter the global propagation properties of the flow, i.e. no
instabilities are introduced as in the magnetic tower case.

\section{Implications for Jet Observations, Experiments and Future Work}
\label{discu}

The results of the simulations help guide our understanding  
 of the evolution of PFD magnetic towers.  In particular the simulations
show new details of the cavity-jet connection, the evolution of the
tower given different assumptions (cooling, rotation, etc.) as well as providing
some insight into the stability properties of the central
jets which form in the flow.

For non-relativistic collimated flows magnetic towers have been
proposed as mechanisms for launching some classes of YSO.  While
the flows downstream at observable distances ($>10^3$\,AU) 
seem to be kinetic energy dominated, at smaller scales a PFD region may be
expected.  As Hartigan et al. (2007) have shown, what few measurements
of magnetic fields exist in YSO jets indicate there must be a region
of sub-Alv\'enic, PFD dominated flow on scales of order 100\,AU or less.  In
addition, these simulations demonstrate (and as laboratory experiments
have shown) the long term evolution of magnetic towers may yield a
series of collimated clumps whose magnetization properties vary over
time.  In this way PFD flows may evolve into kinetic energy dominated
jets at large distances from the central engine.  Planetary Nebula (PN)
offer another potential application of non-relativistic PFD dominated
flows.  Magnetic fields are already expected to play an important
role in launching pre-Planetary Nebulae (PPN) based on an observed
mismatch between momentum in the PPN flow and momentum available
through radiation.  A number of papers have discussed how strong
magnetic fields might create PPN or PN collimated flows (Blackman
et al. 2001a,b, Frank \& Blackman 2004, Matt, Frank \& Blackman 2006).  
Observations of PPN and PN offer morphological
similarities to the kinds of features seen in our simulations, such
as hollow lobes and axial clumps.  Future work might address these
connections.

Of particular importance is the connection between the models
presented in this paper and recent ``laboratory astrophysics''
experiments.  These studies utilized Pulsed Power technologies and
were successful in creating high Mach number, fully radiative,
magnetized outflows (Lebedev et al. 2005, Ciardi et al. 2007,
Ampleford et al. 2008, Ciardi et al. 2009, Frank et al. 2009).
The outflows were created when TW~electrical pulses (1\,MA, 250\,ns) 
are applied to a radial array of fine metallic wires. Lorentz 
forces ablated plasma from the wires creating an ambient plasma 
above the array. After the complete ablation of wires near the 
central electrode, the current switches to the plasma and creates a 
magnetic cavity with a central jet (i.e. a magnetic tower). The 
central part of the jet is confined and accelerated by the 
pressure of the toroidal field. Return current flows along the 
walls of the magnetic cavity, which is in turn confined by the 
thermal pressure and by inertia of the ambient plasma. As the
magnetic cavity expands, the jet becomes detached and propagates 
away from the source at $\sim$200\,km\,s$^{-1}$.  Instabilities which resemble 
the kink mode ($m =\,$1) develop within the body of these jets 
fragmenting them into well collimated structures with characteristic 
axial non-uniformities.

Thus the evolution of magnetic towers in the laboratory show a range
of features that are strikingly similar to what is seen in our
simulations. This concordance is all the more noteworthy in that
our initial conditions were in no way tuned to the experiments
and are, in fact, a modified version of what can be found in a number
of purely astrophysical studies \citep[e.g.][]{li06}. 
Thus it appears that the laboratory
experiments and the simulations support each other, as well as the
conclusion that both are revealing generic properties of PFD outflows.

While we did not study relativistic outflows, some aspects of the
comparative  behavior  between HD and PFD jets revealed by our
models might still apply. The fragmentation of the PFD magnetic
tower core, for example, implies that rather than continuous jet
beams we would expect high resolution observations to reveal
essentially ``clumpy'' jets with a distribution of velocities,
densities and magnetization.  In this way our models,
except the cooling ones,
can be considered
to articulate classes of flow features in AGN radio jets
\citep{tavecchio03}, X-ray binaries \citep{xray} and, perhaps, GRBs
\citep{grb}. But 
both
relativistic 
and different radiative cooling
generalizations are needed to confirm
or refute the implications of our present calculations  for such
regimes.

There is opportunity for future work to focus more closely on the links with the laboratory
experiments.  In particular, issues related to the development of
kink mode instabilities, their non-linear resolution and the evolution
of clumpy magnetized jets should be explored more fully and in more
detail. 

Regarding the effect of rotation at the base of the jets on
their stability, we note that \citet{moll} have carried out 3-D
simulations of magnetocentrifugally driven, conical jets, and found
that kink instabilities are stronger when a rigid rotation profile
is imposed, in comparison to a Keplerian rotation profile.  Rigid
rotation seems to induce a shearless magnetic field \citep{moll}.
A direct comparison with our calculations must be made carefully
though; our initial magnetic configuration has a dominant toroidal
component and no radial component, while the initial field setup
of \citet{moll} is purely radial. Also, our rotating magnetic tower
is continually affected by injection of magnetic 
flux, which is not the case of the conical jets of \citet{moll}.

\section{CONCLUSIONS}
\label{conclu}

We have carried out 3-D ideal MHD simulations of PFD and HD dominated jets to 
compare their structure and evolution subject to the same injected energy flux, and to 
study the effects of cooling and jet rotation on the jet stability. 
We note that our HD cases can, in principle, emulate asymptotic propagation
regimes of magneto-centrifugally launched jets if those jets become kinetic energy dominated 
at large distances.  Magnetic towers will however remain PFD at large radii.

Our simulations read us to the following conclusions.  Helical
localized magnetic fields injected into a region of low pressure
will launch PFD, magnetic towers via magnetic pressure gradients.
Towers consist of a low density low beta plasma, the radial
collimation of which is caused by the pressure of the external plasma. Within
the towers a higher density, higher beta jet forms collimated by the
magnetic field lines located within the cavity.

We found that PFD jets create structures that are more susceptible
to instabilities relative to purely hydrodynamical jets given the same injected
energy flux.  Unstable modes in the magnetic towers
differ according to conditions within the flow. The adiabatic PFD
jet is unstable to  $m=\,$0 and~2 mode perturbations, and its core
adopts a elliptical cross-section.  On the other hand, the PFD jet
with a Keplerian rotating base exhibits an $m=\,$1 kink mode
instability.  The beam is not completely destroyed but adopts a
chaotic clumpy structure.  Base rotation causes a slow amplification
of the toroidal field exacerbating a pressure unbalance in the jet's core
that leads to instability. The cooling PFD jet also shows a
$m=\,$1 kink mode instability.  Cooling reduces the thermal energy
of the jet's core, making the thermal pressure insufficient to damp
the magnetic pressure kink perturbations. The cooling PFD beam shows
the fastest growth rate of the kink instability.

Our magnetic tower (PDF jets) simulations are in good agreement with the
laboratory experiments of \citet{lab1}.  In both our simulations and the experiments: 
(1) jets carry axial
currents which return along the contact discontinuities, (2) the jet
cores have a high $\beta$, (3) jet beams and cavities are PFD, (4) jets are eventually corrugated by
current driven instabilities becoming a collimated chain of magnetized ``clumps'' or ``bullets''.  The
similarity between our models and the experiments is particularly noteworthy
 because our implementation was not tuned to
represent the laboratory results.  This strengthens the case for the usefulness 
of laboratory experiments in articulating new features of astrophysical MHD 
flows in cases where similarity conditions can be obtained. 

We found that PFD jets decelerate by about 20\% relative to the HD
ones given the same injected energy flux. This is because PFD jets
produce not only axial but radial expansion due to magnetic pressure.
All of the pre-collimated energy flux of the HD case is more
efficiently directed to axial mechanical power.  Also, the long
term evolution of PFD jets yield a series of collimated clumps, the
magnetization properties of which may vary over time.  PFD flows
may thus eventually evolve into HD jets at large distances from the
central engine.

Our work shows that outflows launched as magnetic towers show a
different behavior compared with those launched by magneto-centrifugal
(MCL) mechanisms when the MCL flows become asymptotically kinetic
energy flux dominated. As it was shown by Hartigan et al. 2007, in
YSO flows some mechanism may be needed to reduce the magnetization
of plasma close to the jet source.  If these flows begin as magnetic
towers then the disruption of the central jets via kink modes may
provide a means to produce collimated high beta clumps of material
as is observed in HH flows.  Thus our work may help to lead methods
for distinguishing between different launch mechanisms by providing
descriptions of asymptotic flow characteristics where observations
might be possible.

\acknowledgments
Financial support for this project was provided by the Space Telescope
Science Institute grants HST-AR-11251.01-A and HST-AR-12128.01-A;
by the National Science Foundation under award AST-0807363; by the
Department of Energy under award DE-SC0001063; and by Cornell
University grant 41843-7012. SL acknowledges support from 
EPSRC Grant No.~EP/G001324/1. We thank David Meier and Neil Turner 
for helpful discussions.


\begin{thebibliography}{}
\bibitem[Ampleford et al.(2008)]{ampleford08} Ampleford, D.~J., 
Lebedev, S.~V., Ciardi, A., et al., 2008, Physical Review Letters, 100, 
035001 
\bibitem[Bacciotti et~al.(200)]{bacciotti}
Bacciotti, F., Ray, T.P., Mundt, R., Eisffoel, J., 
\& Solf, J., 2002, ApJ, 576, 222
\bibitem[Blackman et al.(2001)]{blackman01} 
Blackman, E.~G., Frank, A., \& Welch, C. 2001, ApJ, 546, 288 
\bibitem[Blackman(2007)]{blackman07}
Blackman E.G., 2007, Ap\&SS, 307, 7 
\bibitem[Blandford \& Payne(1982)]{blandford82} 
Blandford, R.~D., \& Payne, D.~G. 1982, MNRAS, 199, 883
\bibitem[Blondin et al.(1990)]{blondin90} 
Blondin, J.~M., Fryxell, B.~A., \& Konigl, A. 1990, ApJ, 360, 370 
\bibitem[Boyd \& Sanderson(2003)]{book} 
Boyd, T.~J.~M., Sanderson, J.~J., 2003, The Physics of Plasmas,
\bibitem[Carrasco-Gonz{\'a}lez et al.(2010)]{carrasco10} 
Carrasco-Gonz{\'a}lez, C., Rodr{\'{\i}}guez, L.~F., Anglada, G., 
\bibitem[Carroll-Nellenback et al.(2011)]{bear2} 
Carroll-Nellenback, J.~J., Shroyer, B., Frank, A., \& Ding, C., 2011, 
arXiv:1112.1710
\bibitem[Ciardi et al.(2007)]{lab2} 
Ciardi, A., et al. 2007, Phys. of Plasmas, 14, 056501 
\bibitem[Ciardi et al.(2009)]{ciardi9}
Ciardi, A., Lebedev, S.~V., Frank, A., et al., 2011, ApJL, 691, L147
\bibitem[Cunningham et al.(2009)]{astrobear} 
Cunningham A.~J., Frank A., Varni{\`e}re P., Mitran S., 
\& Jones, T.~W. 2009, ApJS, 182, 519 
\bibitem[Dalgarno \& McCray(1972)]{dm} 
Dalgarno A., McCray R.~A. 1972, ARA\&A, 10, 375
\bibitem[Frank et al.(1998)]{frank98} 
Frank, A., Ryu, D., Jones, T.~W., \& Noriega-Crespo, A. 1998, 
ApJL, 494, L79 
\bibitem[Frank et al.(2009)]{frank09} Frank, A., Ciardi, A., 
Yirak, K., \& Lebedev, S., 2009, Revista Mexicana de Astronom\'{\i}a 
y Astrof\'{\i}sica, Conference Series, 36, 193
\bibitem[Gardiner \& Stone(2008)]{CT} 
Gardiner, T.~A., Stone, J.~M., 2008, Journal of Computational 
Physics, 227, 4123
\bibitem[Hardee \& Stone(1997)]{hardee97} 
Hardee, P.~E., \& Stone, J.~M. 1997, ApJ, 483, 121 
\bibitem[Huarte-Espinosa et al.(2011)]{huarte11} 
Huarte-Espinosa, M., Frank, A., Blackman, E., 2011, IAU Symposium, 275, 87
\bibitem[Kato et al.(2004)]{kato} 
Kato, Y., Hayashi, M.~R., Matsumoto, R., 2004, \apj, 600, 338
\bibitem[Konigl \& Pudritz(2000)]{konigl00} 
Konigl, A., Pudritz, R.~E., 2000, Protostars and Planets IV, 759
\bibitem[Krasnopolsky et al.(1999)]{krasnopolsky99} Krasnopolsky, R., 
Li, Z.-Y., \& Blandford, R., 1999, \apj, 526, 631
\bibitem[Krasnopolsky et al.(2003)]{krasnopolsky03} Krasnopolsky, R., 
Li, Z.-Y., \& Blandford, R.~D., 2003, \apj, 595, 631
\bibitem[Kruskal et al.(1958)]{kruskal} Kruskal, M.~D., 
Johnson, J.~L., Gottlieb, M.~B., Goldman, L.~M., 1958, Physics of Fluids, 1, 421 
%
\bibitem[Lery \& Frank(2000)]{lery00} Lery, T., \& Frank, A., 
2000, \apj, 533, 897
%
\bibitem[Lebedev et al.(2005)]{lab1} 
Lebedev, S.~V., et al.(2005), MNRAS, 361, 97
\bibitem[Li et al.(2006)]{li06} 
Li, H., Lapenta, G., Finn, J.~M., Li, S., \& Colgate, S.~A. 2006, 
ApJ, 643, 92
\bibitem[Lind et al.(1989)]{lind89} Lind, K.~R., Payne, D.~G., 
Meier, D.~L., \& Blandford, R.~D., 1989, \apj, 344, 89
\bibitem[Livio(2004)]{livio04} Livio, M., 2004, Baltic 
Astronomy, 13, 273
\bibitem[Lovelace et al.(2002)]{lovelace02} 
Lovelace, R.~V.~E., Li, H., Koldoba, A.~V., Ustyugova, G.~V., 
\& Romanova, M.~M. 2002, ApJ, 572, 445 
\bibitem[Lynden-Bell(1996)]{bell96} 
Lynden-Bell, D. 1996, MNRAS, 279, 389 
\bibitem[Lynden-Bell(2003)]{bell03} 
Lynden-Bell, D. 2003, MNRAS, 341, 1360 
\bibitem[Meier et al.(1997)]{meier97} Meier, D.~L., Edgington, 
\bibitem[Miller-Jones et al.(2007)]{xray} 
Miller-Jones, J.~C.~A., Rupen, M.~P., Fender, R.~P., et al., 
2007, \mnras, 375, 1087 
S., Godon, P., Payne, D.~G., \& Lind, K.~R., 1997, \nat, 388, 350
\bibitem[Mohamed \& Podsiadlowski(2007)]{mohamed07} 
Mohamed S., Podsiadlowski P. 2007, ASPC, 372, 397 
\bibitem[Moll et al.(2008)]{moll} Moll, R., Spruit, H.~C., 
\& Obergaulinger, M., 2008, \aap, 492, 621
\bibitem[Morsony et al.(2010)]{grb} Morsony, B.~J., 
Lazzati, D., \& Begelman, M.~C., 2010, \apj, 723, 267 
\bibitem[Nakamura \& Meier(2004)]{nakamura04} 
Nakamura, M., \& Meier, D.~L. 2004, ApJ, 617, 123 
\bibitem[Ohsuga et al.(2009)]{ohsuga09} Ohsuga, K., Mineshige, 
S., Mori, M., \& Kato, Y., 2009, \pasj, 61, L7 
\bibitem[Ohsuga \& Mineshige(2011)]{ohsuga11} 
Ohsuga, K., \& Mineshige, S., 2011, \apj, 736, 2 
\bibitem[Ouyed \& Pudritz(1997)]{ouyed97} 
Ouyed, R., \& Pudritz, R.~E. 1997, ApJ, 482, 712 
\bibitem[Pudritz(2004)]{pudritz04} Pudritz, R.~E., 
2004, \apss, 292, 471 
\bibitem[Pudritz et al.(2007)]{pudritz07} 
Pudritz, R.~E., Ouyed, R., Fendt, C., \& Brandenburg, A. 2007, 
Protostars and Planets V, 277 
\bibitem[Reipurth et al.(2002)]{reipurth02} Reipurth, B., 
Heathcote, S., Morse, J., Hartigan, P., Bally, J., 2002, \aj, 123, 362
\bibitem[Shafranov(1958)]{shafranov} Shafranov, V.~D., 1958, 
Soviet Journal of Experimental and Theoretical Physics, 6, 545 
\bibitem[Shibata \& Uchida(1986)]{shibata86} 
Shibata, K., \& Uchida, Y. 1986, PASJ, 38, 631
\bibitem[Song \& Cao(1983)]{song} Song, M.-T., Cao, T.-J., 1983, \caa, 7, 159
\bibitem[Suzuki-Vidal et~al.(2010)]{suzuki} 
Suzuki-Vidal, F., Lebedev, S.~V., Bland, S.~N., et al., 2010, 
IEEE Transactions on Plasma Science, 38, 581
\bibitem[Takeuchi et al.(2010)]{takeuchi10} Takeuchi, S., Ohsuga, 
K., \& Mineshige, S., 2010, \pasj, 62, L43
\bibitem[Tavecchio et al.(2003)]{tavecchio03} Tavecchio,
F., Ghisellini, G., \& Celotti, A., 2003, \aap, 403, 83
\bibitem[Ustyugova et al.(2000)]{ustyugova00} 
Ustyugova, G.~V., Lovelace, R.~V.~E., Romanova, M.~M., Li, H., 
\& Colgate, S.~A. 2000, ApJ, 541, L21
\end{thebibliography}
\end{document}